\documentclass[epj]{svjour}
\usepackage{graphics}
\usepackage{amsmath}
\usepackage{amsfonts}
\usepackage{amssymb}
\usepackage{xcolor}
\newcommand{\angstrom}{\textup{\AA}}

\renewcommand\vec[1]{\mathbf{#1}}
\begin{document}
\title{The essential role of surface pinning in the dynamics of charge density waves submitted to external dc fields}

\author{E. Bellec\inst{1,2}, V.L.R. Jacques\inst{2}, J. Caillaux\inst{2} and D. Le Bolloc'h\inst{2}
}                     
\institute{ESRF – The European Synchrotron, 71 Avenue des Martyrs, 38000 Grenoble, France \and Laboratoire de Physique des Solides, UMR 8502 CNRS, Université Paris-Saclay, 91405 Orsay Cedex, France}
\date{Received: date / Revised version: date}
%
\abstract{
A Charge Density Wave (CDW) submitted to an electric field displays a strong shear deformation because of pinning at the lateral surfaces of the sample. This CDW transverse pinning was recently observed
but has received little attention from a theoretical point of view until now despite important consequences on electrical conductivity properties.
Here, we provide a description of this phenomenon by considering a CDW submitted to an external dc electric field and constrained by boundary conditions including both longitudinal pinning due to electrical contacts and transverse surface pinning. A simple formula for the CDW phase is obtained in 3D by using the Green function and image charges method. In addition, an analytical expression of the threshold field dependence on both length and sample cross section is obtained by considering the phase slip process. We show that the experimental data are well reproduced with this model and that bulk pinning can be neglected. This study shows that the dynamical properties of CDW systems could be mainly driven by boundary effects, despite the comparatively huge sample volumes. 
%
} 
\authorrunning
\titlerunning
\maketitle
\section{Introduction}
\label{intro}

 When a sufficiently large electrical current is applied to a charge density wave (CDW) system, a non-linear current may be induced \cite{Cox2008,RevModPhys.60.1129}. This current has a periodic structure both in time \cite{Onishi_2017} and space \cite{PhysRevLett.100.096403,PhysRevB.85.035113} and is believed to be due to a collective transport of charge based on a traveling periodic array of topological $2\pi$ solitons \cite{PhysRevB.94.201120}. The CDW can be described as an elastic medium evolving under the application of an external electric field \cite{Feinberg}. When this field exceeds a threshold value $E_{th}$, the CDW periodicity is broken and a soliton is created reducing the total elastic energy. This phase-slip mechanism \cite{GORKOV1983,gor1984generation} has been widely discussed in the literature as a thermally activated nucleation \cite{PhysRevLett.68.2066,Gill_1986} or as a quantum process \cite{PhysRevB.48.4860,MAKI1995313}.

The sliding CDW phenomenon was extensively studied as a function of temperature \cite{PhysRevB.85.241104}, under a continuous laser photo-illumination \cite{PhysRevB.70.075111} or as a depinning process induced by a short laser pulse excitation \cite{PhysRevLett.117.156401}. In all these phenomena, CDW pinning plays a fundamental role. The aim of this paper is to theoretically treat pinning effects as a whole including pinning at lateral surfaces when the CDW is submitted to an applied field and the incidence of this pinning on the threshold electric field $E_{th}$ as a function of sample dimensions.

There are two types of pinning at the sample boundaries. Longitudinal pinning due to the two electrical contacts along the CDW wavevector $2k_F$ has been observed by several techniques. At the contact position, the CDW phase is constant as a function of external current. Resistivity measurements show that the threshold field diverges with decreasing sample lengths in NbSe$_3$ \cite{PhysRevB.32.2621,YETMAN1987201} and in TaS$_3$ \cite{MIHALY1983203}. On the other hand, the consequence of longitudinal pinning under current leads to CDW compression/expansion in the vicinity of the two contacts as reported by x-ray diffraction \cite{PhysRevLett.80.5631,PhysRevB.61.10640} and transport measurements\cite{PhysRevB.57.12781}.

However, longitudinal pinning is not the only constraint applied to the CDW. Resistivity measurements have shown that transverse pinning at lateral sample surfaces may also play an important role since the threshold field $E_{th}$  diverges with decreasing sample cross sections in NbSe$_3$ \cite{PhysRevB.46.4456,YETMAN1987201} and in TaS$_3$ \cite{BORODIN198673}. In a recent paper \cite{PhysRevB.101.125122}, transverse pinning was indeed observed at the local scale by scanning x-ray microdiffraction in NbSe$_3$. The CDW displays a large and continuous transverse deformation over an impressively large distance, spanning from one lateral surface to the other. This represents several tens of micrometers which is 4 orders of magnitude larger than the CDW wavelength. 

From a theoretical point of view, the threshold field $E_{th}$ has been mainly estimated by considering 1D models and the distance $L_x$ between the two electrical contacts. Feinberg and Friedel\cite{Feinberg} proposed a phenomenological relation between $E_{th}$ and $L_x$ by considering CDW bulk impurity pinning. On the other hand, Batistic {\it et al} \cite{Batisti} computed $E_{th}$ by considering longitudinal pinning and found a power law dependence $E_{th}\approx 2.55 L_x^{-\alpha}$ where $\alpha = 1.23 \pm 0.05$. In this 1D model, $E_{th}$ drops to 0 for large $L_x$ in contradiction with experiments showing that $E_{th}$  converges to finite values \cite{PhysRevB.32.2621}. We will show in the following that taking into account lateral surface pinning naturally leads to convergence towards a non-diverging finite threshold. 
On the other hand, Borodin {\it et al}\cite{BORODIN198673} considered the effect of the sample cross section and found a power law behavior $E_{th}\propto A^{-1/2}$ with $A = L_y\times L_z$.

In this article, we calculate $E_{th}(L_x,L_y,L_z)$  by taking into account both longitudinal {\it and} transverse pinning in the known 3D free energy\cite{PhysRevB.48.4860,hayashi2000ginzburglandau} by considering the phase-slip process and using the Green function and image charges methods.

\section{Behavior of a CDW under electric field with contact and surface pinning}
\label{sec:1}

\subsection{CDW phase equation, Green function and image charges method}
\label{sec:1_1}

A CDW is described through its periodic charge modulation $\rho(\vec{r})=A(\vec{r})\cos(2k_F x+\phi(\vec{r}))$ where $A(\vec{r})$ and $\phi(\vec{r})$ are respectively the CDW amplitude and phase with spatial dependence. In this paper, we only consider the evolution of $\phi$ under an applied electric field with component $E$ along the $x$ direction. The CDW behavior can be described by the following free energy considering only the phase variations: 

\begin{equation}\label{FreeEnergyPhase3D} 
\begin{split}
\mathcal{F}[\phi] \propto \int_\mathcal{V} d^3\vec{r} \, \biggl\{ C^{ij}\phi_i\phi_j+V_{imp}(\phi) -\eta Ex\phi_x\biggr\}
\end{split}
\end{equation}
\begin{equation*}
\phi_i = \partial_i \phi, \qquad\vec{r}\in \mathcal{V} \equiv \left\lbrace \vec{r}\in \mathbb{R}^3\biggl \lvert \lvert x_i \rvert \leq L_i/2 \right\rbrace
\end{equation*}
where $i,j=x,y,z$, $C^{ij} = c_i c_j \delta^{ij}$ with $c_x,c_y,c_z$ being the CDW longitudinal and transverse elastic coefficients, $L_x$ is the contact distance, $L_y$ and $L_z$ are the transverse sample lengths. We choose to consider a bulk impurity pinning potential $V_{imp}(\phi) \equiv \omega_0^2[1-\cos(\phi)]$ with the pinning frequency $\omega_0$ often used in the literature \cite{PhysRevB.17.535,gruner2018density}. We will show in the following that this term can be neglected compared to the surface pinning effect. Finally, the last term corresponds to the interaction between the CDW and the applied electric field coupling the longitudinal gradient $\phi_x$ and the applied electric potential $Ex$ where $\eta$ is a temperature dependent coupling coefficient \cite{hayashi2000ginzburglandau}.

While a randomly distributed bulk impurity pinning can break the CDW long range order \cite{PhysRevB.17.535,lee1979electric} x-ray diffraction experiment in NbSe$_3$ \cite{PhysRevB.101.125122} showed a continuous CDW deformation over tens of micrometers. Furthermore, as measured in another x-ray diffraction experiment \cite{PhysRevLett.80.5631}, the CDW wavevector variations are less than  $0.25.10^{-4}b^*$ below $E_{th}$, hence one can consider $\phi\ll 2\pi$ in the following and use a Taylor development of $V_{imp}(\phi)\approx \omega_0^2\phi^2$. Above $E_{th}$, $2\pi$ solitons nucleate at the electrical contact, hence the phase solution is only correct in the linear current regime below $E_{th}$ which is the regime considered here.
 
Minimizing $\mathcal{F}[\phi]$ with the Euler Lagrange equation $\partial_i\frac{\partial \mathcal{F}}{\partial \phi_i}-\frac{\partial\mathcal{F}}{\partial \phi}=0 $ we obtain:
\begin{equation}\label{phaseEq}
2\left(c_x^2\phi_{xx}+c_y^2\phi_{yy}+c_z^2\phi_{zz}\right)-\omega_0^2\phi \approx \eta E
\end{equation}

The CDW longitudinal pinning at the two electrical contacts is taken into account by setting the phase at zero at $x=\pm \frac{L_x}{2}$\cite{Feinberg}. In addition, the transverse pinning observed in \cite{PhysRevB.101.125122} is included in the model by considering  boundary conditions at the lateral surfaces $y=\pm \frac{L_y}{2}$ and $z=\pm \frac{L_z}{2}$. The two types of pinning leads to the following constraints:
\begin{equation}
\label{boundariesConditions}
\phi(\vec{r}) = 0,\qquad \forall \vec{r} \in \partial \mathcal V
\end{equation}

By rescaling $x_j=c_j\sqrt{\frac{2}{\eta}}x'_j$, $L_j=c_j\sqrt{\frac{2}{\eta}}L'_j$ and $\omega^2=\frac{\omega_0^2}{\eta}$, the phase equation Eq.\ref{phaseEq} and the boundary conditions Eq.\ref{boundariesConditions} become:

\begin{equation}\label{equationChangeVariable} 
\left(\Delta'-\omega^2\right)\phi = E 
\end{equation} 
with the Dirichlet conditions:
\begin{equation}\label{conditionsChangeVariable} 
\phi(\vec{r'}) = 0,\qquad \forall \vec{r'} \in \partial \mathcal V'
\end{equation} 
where $\Delta' = \frac{\partial^2}{\partial x'^2}+\frac{\partial^2}{\partial y'^2}+\frac{\partial^2}{\partial z'^2}$ is the rescaled laplacian operator. 

Eq.\ref{equationChangeVariable} is the screened Poisson equation that satisfies the uniqueness theorem as shown in appendix \ref{appendix:uniqueness}. Therefore,  a solution to this differential equation can be computed by using the Green function and image charge method \cite{riley2002mathematical}. This technique is well established to solve electrostatic problems with boundary conditions. In our case, the source term is $+E$ inside the sample volume. The Green function satisfying the equation $\left(\Delta-\omega^2\right)G(\vec{r}-\vec{r}')=\delta(\vec{r}-\vec{r}')$ has a simple lorentzian expression in Fourier space:
\begin{equation}\label{Green}
G(\vec{q}) = \frac{-1}{\lvert\vec{q}\rvert^2+\omega^2}
\end{equation}

There is no simple analytic solution fulfilling Eq.\ref{equationChangeVariable} with the boundary conditions Eq.\ref{conditionsChangeVariable} in 2D and 3D. We will show later that its resolution requires the use of the Green function and image charges method. As a first step, we describe the image charge construction in 1D before generalizing the procedure to the 2D and 3D cases.

\subsection{The CDW phase in 1D, 2D and 3D pinned by contacts and lateral surfaces}
\label{sec:1_2}

The Eq.\ref{equationChangeVariable} in 1D with $\phi\left(\pm \frac{L_x'}{2}\right)=0$ has an analytical solution:
\begin{equation}\label{1DAnalytic}
\phi_{ana}(x') = \frac{E}{\omega^2}\left[ \frac{\cosh\left(x'\omega\right)}{\cosh\left(\frac{L_x'\omega}{2}\right)}-1\right].
\end{equation}
This 1D solution $\phi_{ana}(x)$ will be later compared to the solution obtained from the Green function method. 

A step-by-step construction of the image charge density and the corresponding charge density wave phase calculation are described in appendix \ref{appendix:stepImageCharge}. The phase solution from the Green function and image charge methods reads :

\begin{equation}\label{1DSolutionSum} 
\phi(x') = -\frac{4E}{\pi}\sum\limits_{n=0}^{+\infty}\frac{(-1)^n\cos\left[(2n+1)\pi\frac{x'}{L_x'}\right]}{(2n+1)\left\lbrace \left[ (2n+1)\frac{\pi}{L_x'}\right]^2+\omega^2 \right\rbrace}
\end{equation}

\begin{figure}
\resizebox{0.43\textwidth}{!}{\includegraphics{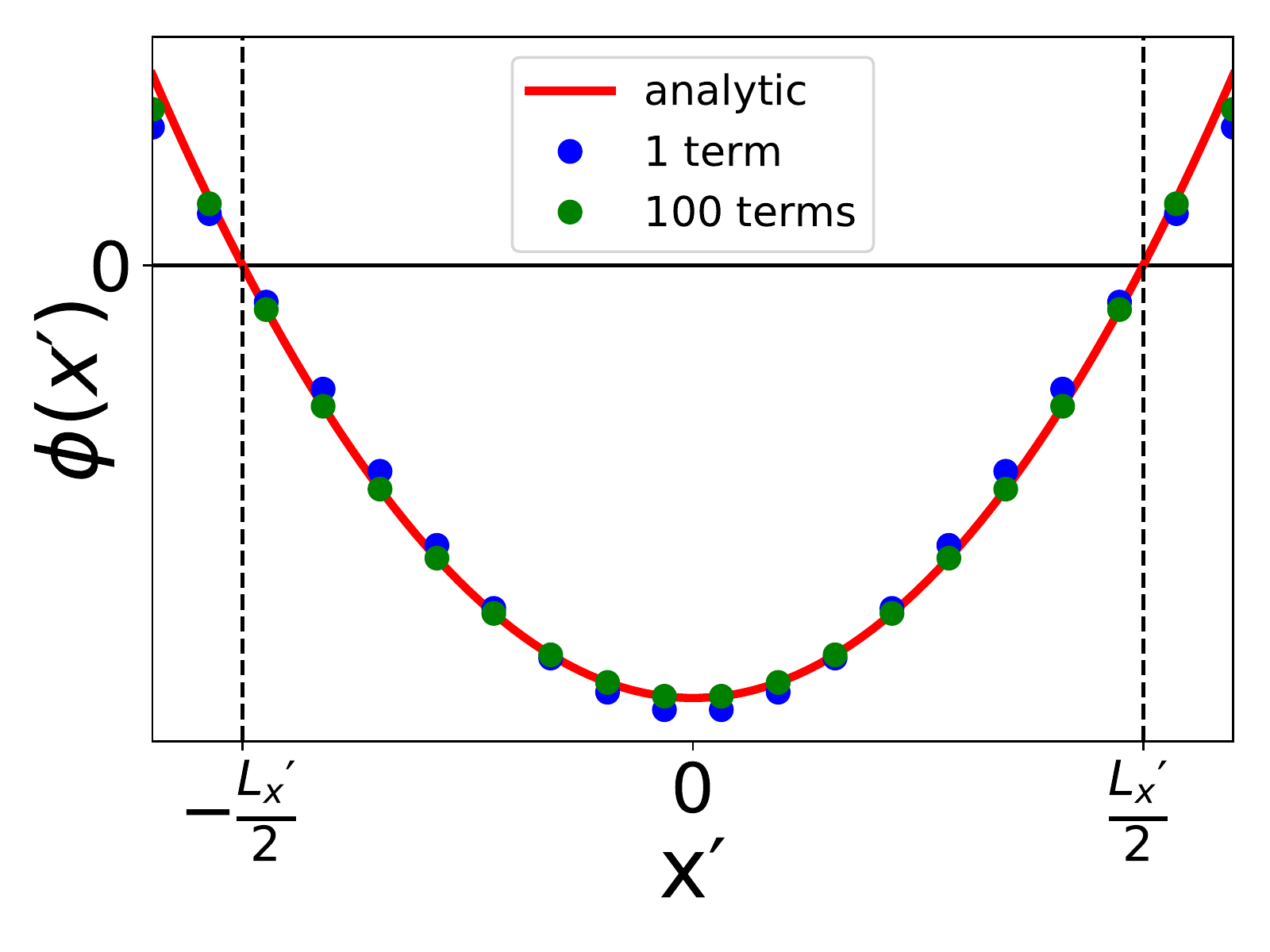}}
\caption{Comparison between the exact solution of the CDW phase $\phi_{ana}$ in 1D and the solution obtained with the Green function and image charges Eq.\ref{1DSolutionSum} with the first term only ($n=0$) and the first 100 terms in the infinite sum (with $E=1$, $L_x'=1$ and $\omega=0$). The  convergence is fast for $\omega<1$, since the first term $n=0$ already gives a solution close to $\phi_{ana}$.}
\label{fig:2}       
\end{figure}

This infinite sum converges quickly with $n$, decreasing as $\frac{1}{n^3}$. The  contribution of the first term ($n=0$) and the first 100 terms ($n$ from 0 to 99) is shown in Fig.\ref{fig:2} and compared to the exact analytic expression Eq.\ref{1DAnalytic}. Note that Eq.\ref{1DAnalytic} and Eq.\ref{1DSolutionSum} start to differ outside the sample boundary $-\frac{L_x'}{2}<x'<\frac{L_x'}{2}$,   which is not surprising since the two solutions have been defined to be valid in the sample limits.

The construction of the image charge density in 2D is similar to the procedure used in 1D and is described in appendix \ref{appendix:stepImageCharge}. The final expression of the 2D CDW phase with contact and transverse surface pinning is the following:

\begin{equation}\label{2DSolutionSum}
\begin{split}
\phi(x',y') = -\frac{16E}{\pi^2}\sum\limits_{n_{x,y}=0}^{+\infty}\frac{(-1)^{n_x+n_y}}{(2n_x+1)(2n_y+1)} \times  \\
\frac{\cos\left[(2n_x+1)\pi\frac{x'}{L'_x}\right]\cos\left[(2n_y+1)\pi\frac{y'}{L'_y}\right] }{ \left[(2n_x+1)\frac{\pi}{L'_x}\right]^2+\left[(2n_y+1)\frac{\pi}{L'_y}\right]^2+\omega^2}
\end{split}
\end{equation}

\begin{figure}
\resizebox{0.5\textwidth}{!}{\includegraphics{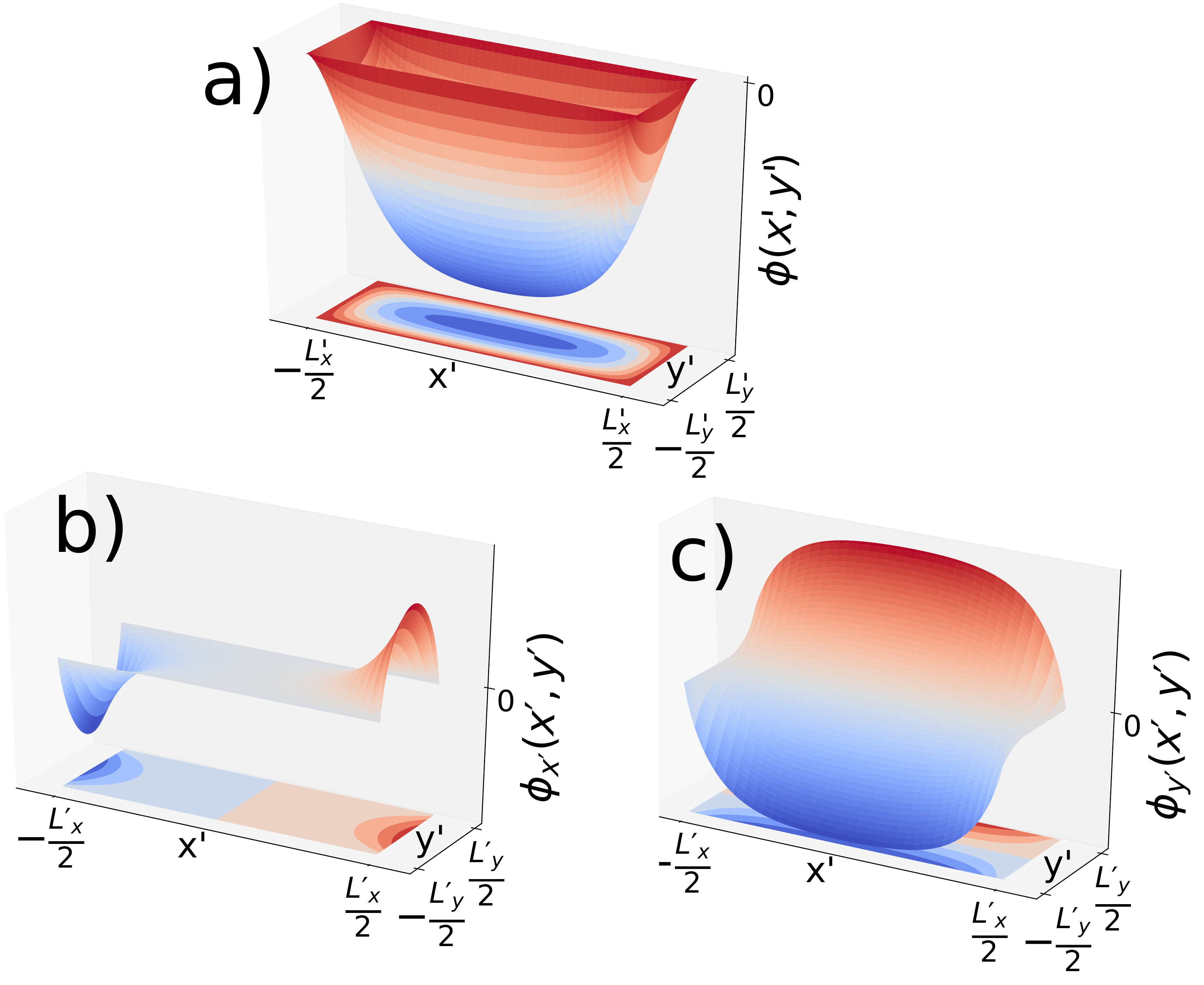}}
\caption{a) The 2D CDW phase obtained from the Green function and image charge method with $\omega = 0$, $E=1$,  $L'_x = 3$, $L'_y=1$ by summing the 100 first terms in $n_x$ and $n_y$. By construction, $\phi=0$ at all sample edges. b) The 2D longitudinal strain, proportional to the phase derivative $\phi_{x'}$, shows two extrema at the two electrical contacts $(x',y')=\left(\pm\frac{ L'_x}{2},0\right)$. c) 2D transverse derivative showing a large shear effect in the middle of the sample.}
\label{fig:3}       
\end{figure}

\begin{figure}
\centering
\resizebox{0.4\textwidth}{!}{\includegraphics{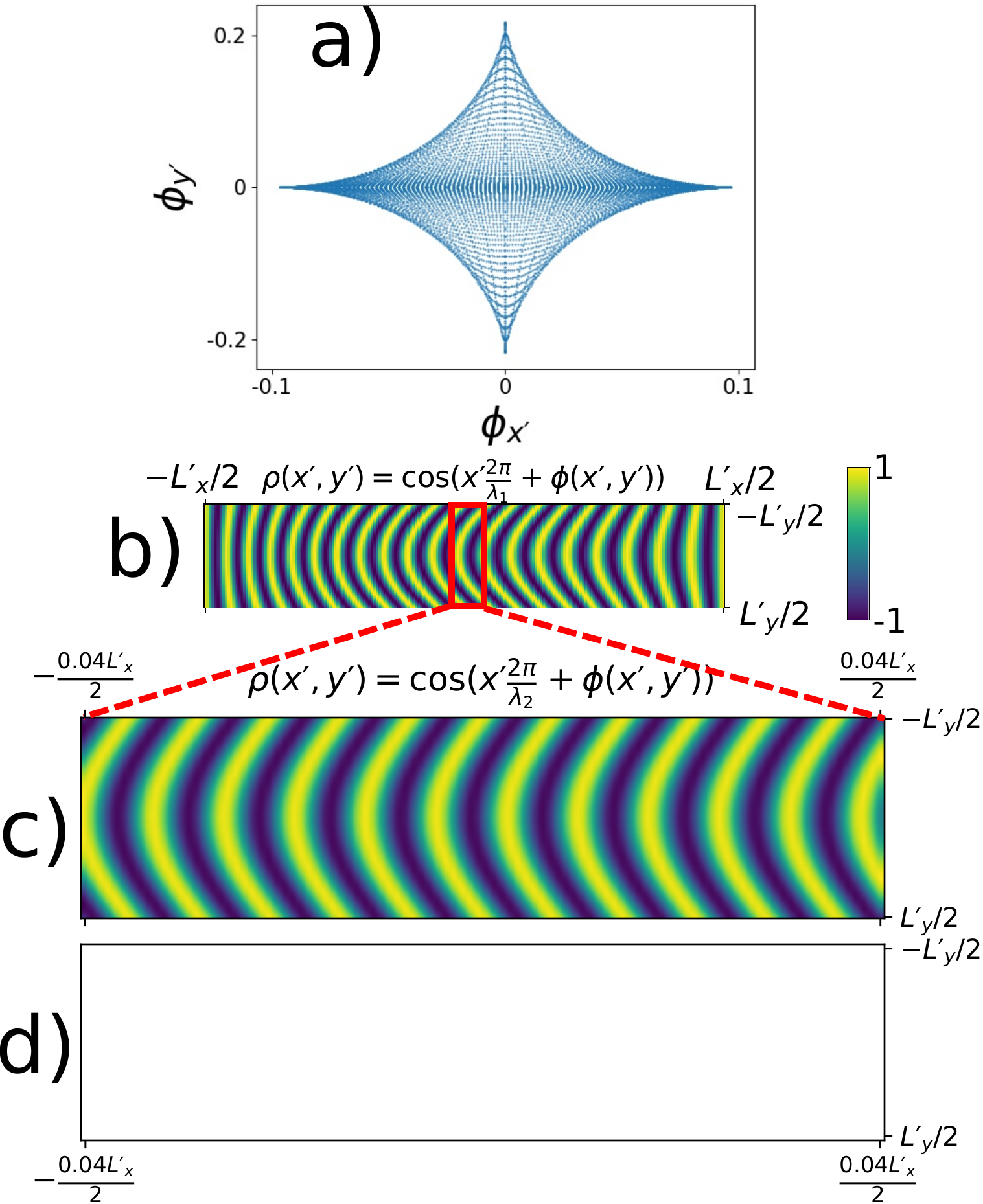}}
\caption{a) Transverse derivative $\phi_{y'}$ 
as a function of the longitudinal one $\phi_{x'}$ obtained from Eq.\ref{2DSolutionSum}. b) The CDW corresponding to $\phi(x',y')$ of Fig.\ref{fig:3}a) displays  a shear effect with a curvature of the wave fronts in the middle part of the sample and a compression-dilation of the CDW period at the two electrical contacts. The CDW wavelength $\lambda$ has been significantly increased for clarity (in reality $\lambda =14\angstrom$ in NbSe$_3$, that is $\lambda \approx 10^{-6}L_x$).  c) Comparison of the theoretical $\rho(x',y')$ with the CDW measured by x-ray diffraction under current from \cite{PhysRevB.101.125122}. Note that we used a different CDW wavelength between b) and c),d)  for a better visualization.}
\label{fig:ComparisonKmap}       
\end{figure}

This 2D CDW phase submitted to an applied field is shown in Fig.\ref{fig:3}a along with its longitudinal (b) and transverse (c) derivatives that are proportional respectively to the longitudinal strain and transverse shear. The corresponding CDW is shown in Fig.\ref{fig:ComparisonKmap}b where we can observe a compression and a dilatation of the CDW period at the two electrical contacts induced by contact pinning. Furthermore, transverse surface pinning induces a shear effect with a curvature of the CDW wave fronts in the middle of the sample.

Note that  the shear is strong when the compression-dilatation is weak and conversely, the curvature is almost zero when the compression and the dilatation are large as can be observed in Fig.\ref{fig:3}b and c. This feature is illustrated in Fig.\ref{fig:ComparisonKmap}a where the transverse phase derivative $\phi_{y'}$ is displayed as function of the longitudinal phase derivative $\phi_{x'}$. To get this figure, $\phi_{x'}$ and $\phi_{y'}$ have been calculated at each point $(x',y')$ of Fig.\ref{fig:3}b and c, and plotted as a 2D graph. From this graph and the 2D phase derivatives represented in Fig.\ref{fig:3}b and c, we observe that $\lvert \phi_{y'}\rvert$ can be large only at position where $\lvert \phi_{x'}\rvert$ is low. This phenomenon is in  agreement with X-ray diffraction experiments showing that  the shear is ten times larger than the longitudinal dilatation-compression in the central part of the sample\cite{PhysRevB.101.125122} and that the longitudinal deformation is larger close to the contacts\cite{PhysRevLett.80.5631,PhysRevB.61.10640}. In Fig.\ref{fig:ComparisonKmap}c and d, we compare the calculated CDW shear in the middle part of the sample with the one measured in \cite{PhysRevB.101.125122}.

As detailed in appendix \ref{appendix:phase-slip}, the phase slip-process occurs at a given threshold field $E_{th}$ when  the strain along $x$ 
($e_{xx}=\frac{\phi_x}{2k_F}$) exceeds a threshold value. The partial derivative $\phi_{x'}$ 
is displayed in Fig.\ref{fig:3}b showing that the longitudinal strain is larger near the contact as observed by several experiments \cite{PhysRevLett.80.5631,PhysRevB.61.10640,PhysRevB.57.12781}. Moreover, the longitudinal strain decreases to zero at the transverse surface $y'=\pm{L'_y}/{2}$. 

From Eq.\ref{1DSolutionSum} in 1D and Eq.\ref{2DSolutionSum} in 2D, one can generalize the solution to the 3D case. The full expression $\phi(\vec{r}';E)$
(where we made explicit that $\phi$ depends on the field value $E$) is given in appendix \ref{appendix:3Dsolution} along with a surface plot of $\phi$ and its longitudinal derivative in Fig.\ref{fig:App4}. As in the 2D case, the longitudinal derivative $\phi_{x'}$ shows a maximum at the electrical contact $(x'=\frac{L'_x}{2}$, $y'=z'=0)$ and a minimum on the other side $(x'=-\frac{L'_x}{2}$, $y'=z'=0)$. 

Since $\phi(\vec{r})$  is based on a fast convergent sum, one can only keep the first term of the 3D expression:
\begin{equation}
\phi(\vec{r})\approx-\frac{32}{\pi^5}E\eta \beta \cos(\pi \frac{x}{L_x})\cos(\pi \frac{y}{L_y})\cos(\pi \frac{z}{L_z})
\end{equation}
with
$$\beta= \frac{1}{\frac{c_x^2}{L_x^2}+\frac{c_y^2}{L_y^2}+\frac{c_z^2}{L_z^2}+\frac{\omega_0^2}{2\pi^2}}$$
At the first order, the CDW phase $\phi(\vec{r})$ is proportional to a simple product of cosine functions. Considering the first term only is a good approximation in the case $\frac{L_x}{c_x}\sim \frac{L_y}{c_y}\sim \frac{L_z}{c_z}$ and $\omega_0 \lesssim 1$ (see the 1D case in Fig.\ref{fig:2}).

\section{The threshold electric field $E_{th}$}
\label{sec:3}

\subsection{$E_{th}$ as a function of electrical contact distance}
\label{sec:3_1}

As described in appendix \ref{appendix:phase-slip}, a soliton nucleates spontaneously whenever $\phi_{x}$ 
exceeds a threshold value $\phi'_c$. Therefore, the threshold electric field $E_{th}$ is defined by: 

\begin{equation}\label{phaseSlipEq}
 \phi_x\left(\frac{L_x}{2},0,0;E_{th}\right) =  \phi'_c
\end{equation}
Taking into account the derivative of  $\phi(x',y',z')$ in 3D (Eq.\ref{3DSolutionSum} in appendix \ref{appendix:3Dsolution}) where the longitudinal strain is maximum at $(x',y',z')=\left(\frac{L'_x}{2},0,0\right)$. Making the reverse change of variable (Eq.\ref{equationChangeVariable} and Eq.\ref{conditionsChangeVariable}) and using the following relation:
\begin{equation}
\sum\limits_{n_x=0}^{\infty}\frac{1}{(2n_x+1)^2+a^2} = \frac{\pi}{4a}\text{tanh}\left(\frac{\pi a}{2}\right), 
\end{equation}
we obtain the threshold field $E_{th}$ versus the sample dimensions $L_x$, $L_y$,$L_z$ and the bulk impurity pinning $\omega_0$:

\begin{equation}\label{Eth}
E_{th} = \frac{\phi_c'\pi^3c_x^2}{8\eta L_x\sum\limits_{n_{y,z}=0}^{+\infty} \frac{(-1)^{n_y+n_z}}{(2n_y+1)(2n_z+1)a_{n_y,n_z}} \text{tanh}\left(\frac{\pi a_{n_y,n_z}}{2}\right)}
\end{equation}
where we defined for ease of notation:
\begin{equation}\label{anynz}
\begin{split}
a_{n_y,n_z}=\frac{L_x}{c_x}\left(\left[(2n_y+1)\frac{c_y}{L_y}\right]^2+\left[(2n_z+1)\frac{c_z}{L_z}\right]^2 \right.\\
\left. +\frac{1}{2}\left(\frac{\omega_0}{\pi}\right)^2 \right)^{1/2}
\end{split}
\end{equation}

Since $a_{n_y,n_z}$ is proportional to $L_x$, one can easily find the two limits. For small electrical contact distances $L_x\rightarrow 0$, using $\sum_{n=0}^\infty\frac{(-1)^n}{2n+1}=\frac{\pi}{4}$ we find

$$
\lim\limits_{L_x \ll \gamma}  E_{th}=\frac{4\phi_c' c_x^2}{\eta} \frac {1}{L_x},
$$

where 

$$
\gamma = \frac{2c_x}{\pi\sqrt{\left(\frac{c_y}{L_y}\right)^2+\left(\frac{c_z}{L_z}\right)^2+\frac{1}{2}\left(\frac{\omega_0}{\pi}\right)^2}}
$$

In this case, the threshold diverges as $1/L_x$ in agreement with the experiments as we will see later.
On the contrary, the threshold for longer samples:

$$\lim\limits_{L_x \gg \gamma}E_{th}=  \frac{\phi_c'\pi^3c_x}{8\eta}\frac{1}{S(L_y,L_z)}$$
with
\begin{equation}
\begin{split}
S(L_y,L_z)=\sum\limits_{n_{y,z}=0}^{+\infty} \frac{(-1)^{n_y+n_z}}{(2n_y+1)(2n_z+1)} \\
\times \frac{1}{\sqrt{\left[(2n_y+1)\frac{c_y}{L_y}\right]^2+\left[(2n_z+1)\frac{c_z}{L_z}\right]^2+\left(\frac{\omega_0}{2\pi}\right)^2}}
\end{split}
\end{equation}

does not depend on L$_x$ anymore. The threshold remains constant above a given contact distance even if $\omega_0=0$. This last point is crucial, because it shows that the experimentally observed saturation of $E_{th}$ for long samples is naturally reproduced by considering only surface and contact pinning, with no need of bulk pinning. 

Eq.\ref{Eth} has been used to fit several transport measurements reporting the dependence of $E_{th}$ on sample length in NbSe$_3$ and TaS$_3$\cite{PhysRevB.32.2621,PhysRevB.29.755,MIHALY1983203}.  Eq.\ref{Eth} contains 6 free parameters ($c_x,c_y, c_z,\omega_0, \eta$ and $\phi'$) assuming the crystal dimensions $(L_x,L_y, L_z)$ are known. Nevertheless, this too large number of free parameters can be significantly reduced to the 4 following parameters : 
$$
\left\{
\begin{array}{ll}
      p_1 =&\frac{ \phi'_c\pi^3c_x^2}{8\eta}\\
      p_2 =&\frac{c_y}{L_yc_x} \\
      p_3 =&\frac{c_z}{L_zc_x} \\
      p_4=&\frac{\omega_0}{\sqrt{2}\pi c_x}\\
\end{array} 
\right.
$$
The expression of $E_{th}$ with those parameters is given in appendix \ref{fitting}.

This number can be again reduced by making several assumptions.
 Despite the lack of data about the phason mode in NbSe$_3$, the phason dispersion curve has been measured in K$_{0.3}$MoO$_3$ \cite{PhysRevB.43.8421} showing that the two transverse elastic constants are similar. We assume that it also the case in NbSe$_3$ and set the constraint $c_y=c_z$.

Furthermore,  standard NbSe$_3$ and TaS$_3$ crystals display very elongated shapes, few millimeters long, tens of micrometers wide but only few micrometers thick.  We thus assume $L_y\gg L_z$ ($p_2\ll p_3$) and set $p_2=0$ in the fit since the two contributions  add up in the formula.

Furthermore, the bulk impurity frequency pinning $\omega_0$ is a phenomenological parameter introduced in previous models\cite{Feinberg} to explain why the threshold does not tend towards zero for large $L_x$ distances. However, the impurity pinning $\omega_0$ does not bring any crucial information in our model since the finite limit for large $L_x$ is naturally obtained by the lateral surface pinning introduced here (see appendix \ref{fitting}). Therefore, impurity pinning frequency $\omega_0$ is set to zero in the following.

The evolution of $E_{th}$ as a function of $L_x$ taken from ref\cite{PhysRevB.32.2621} has been correctly fitted for different sets of free parameters (see Fig.\ref{fig:4} in the appendix \ref{fitting}). 
The overall $E_{th}$ profile including its convergence towards a  constant value for large $L_x$ is well reproduced with $\omega_0=0$ which confirms that the physical properties of sliding CDW is mainly driven by surface pinning effects and not by bulk pinning.

\begin{figure}
\resizebox{0.42\textwidth}{!}{\includegraphics{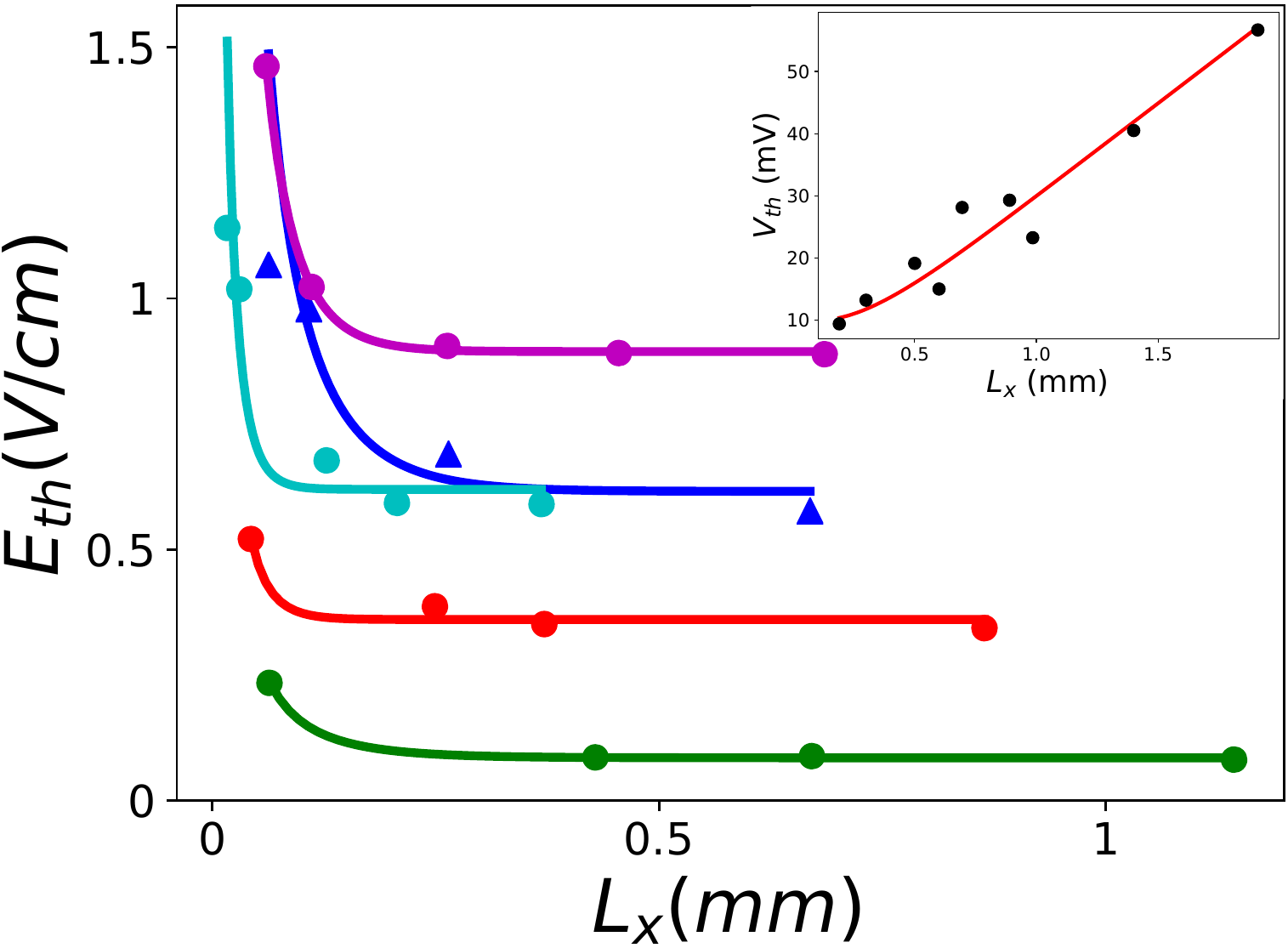}}
\caption{Threshold $E_{th}$ with $L_x$ fitted by using our model including transverse pinning (Eq.\ref{EthFitLx} in appendix \ref{fitting} using $\{p_1,p_2=0,p_3,p_4=0\})$. The experimental dots are taken from \cite{PhysRevB.32.2621} measured in NbSe$_3$ and the blue triangles  from \cite{MIHALY1983203} in TaS$_3$. Inset : Fit of $V_{th}=L_x\times E_{th}$ using Eq.\ref{EthFitLx} of resistivity data from \cite{PhysRevB.29.755}.}
\label{fig:5}       
\end{figure}

Within those assumptions, Eq\ref{Eth} correctly fits several measurements performed in different systems as  in TaS$_3$\cite{MIHALY1983203} or in NbSe$_3$\cite{PhysRevB.32.2621} as shown in Fig.\ref{fig:5} despite our constraints $p_2=p_4=0$, showing that the phase-slip process combined with the transverse surface pinning can explain the non-linear CDW current measured in these materials.
Finally, the electrical potential $V_{th}=L_x\times E_{th}$ as a function of $L_x$ taken from \cite{PhysRevB.29.755} is also fitted in the inset of Fig.\ref{fig:5}. The model also confirms the linear behavior of $V_{th}$ for large contact distance.

A more systematic study knowing the exact sample dimensions ($L_y$ and $L_z$) would allow us to directly obtain the $c_x/c_y$ and $c_x/c_z$ ratios from resistivity measurements  (the sample transverse dimensions  are sometimes missing in the published data). 

\subsection{$E_{th}$ as a function of the sample cross section}
\label{sec:3_2}

Let's now study the effect of the sample cross section on $E_{th}$ using our model. It is interesting to consider this dependence since the observed increase of $E_{th}$ with decreasing cross sections can not be explained by the bulk pinning frequency $\omega_0$ alone. 
We use data from \cite{BORODIN198673} of the threshold field  as a function of the sample cross section $A = L_yL_z$ in small o-TaS$_3$ samples. The first $(2n_y+1)\frac{c_y}{L_y}$ term in the expression of $a_{n_y,n_z}$ Eq\ref{anynz} can be neglected since $L_y\approx 10L_z\gg L_z$ ($A\approx10L_z^2)$. As in the previous section, we assume $c_y\approx c_z$ and the bulk impurity pinning is neglected ($\omega_0=0$). It remains two free parameters are thus remaining, namely $\left\lbrace m_1=\frac{\phi'_c\pi^3c_x^2} {8\eta L_x},m_2\approx\frac{c_yL_x\sqrt{10}}{c_x} \right\rbrace$ and the fitting function reads:
\begin{equation}\label{EthFitA}
\begin{split}
E_{th,fit \, A}(A,\{m_1,m_2\}) = m_1\times \\
\frac{1}{ \sum\limits_{n_{y,z}=0}^{+\infty} \frac{(-1)^{n_y+n_z}}{(2n_y+1)(2n_z+1)a_{n_y,n_z}} \text{tanh}\left(\frac{\pi a_{n_y,n_z}}{2}\right)}
\end{split}
\end{equation}
with 
\begin{equation}
a^{fit \, A}_{ny,n_z}=m_2\frac{(2n_z+1)}{\sqrt{A}}
\end{equation}

\begin{figure}
\resizebox{0.45\textwidth}{!}{\includegraphics{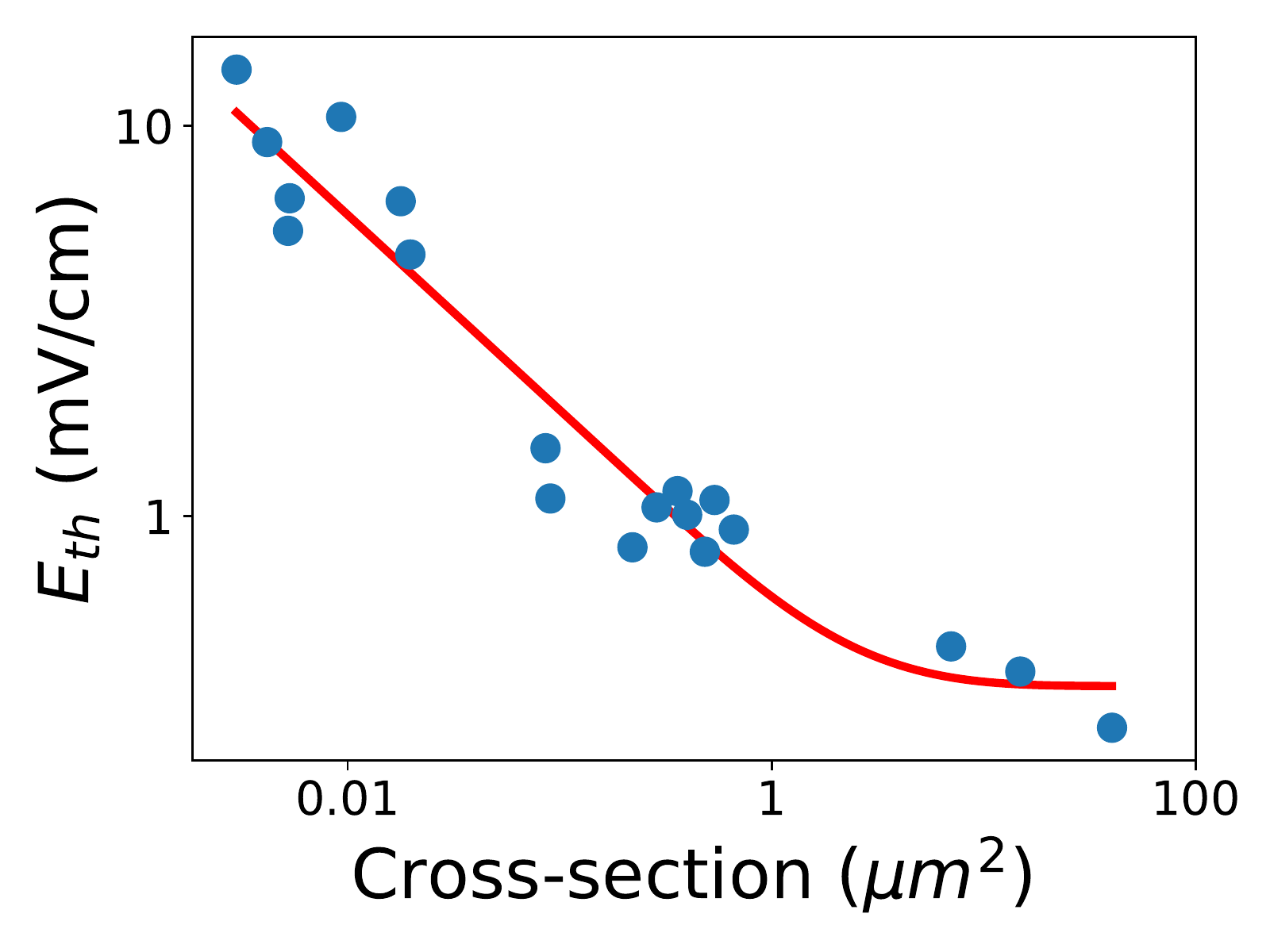}}
\caption{The experimental threshold $E_{th}$ (blue dots) versus the sample cross section in small o-TaS$_3$ samples from \cite{BORODIN198673}. The fit using Eq.\ref{EthFitA} (red line) correctly reproduces the whole behavior especially the increase of $E_{th}$ for decreasing cross sections and the asymptotic constant value for large cross sections (no bulk pinning is used in the fit).}
\label{fig:6}       
\end{figure}

Our model with $\omega_0=0$, $c_y=c_z$ and  $Ly\gg L_z$ correctly fits the experimental data (see Fig.\ref{fig:6}) especially the two extreme cases, the threshold increase for small cross sections and the finite asymptotic value for large cross sections.

The increase of $E_{th}$ for decreasing cross sections can not be reproduced by bulk impurity pinning but is correctly described by considering only surface pinning. 
This adjustment with cross sections is the best illustration that the pinning volume plays little, if any role, in the CDW deformation under current.

It can also be noted that the critical strain is reached at larger fields for smaller cross sections. Due to surface pinning, part of the elastic energy induced by the deformation goes into the transverse shear, thus the longitudinal strain becomes smaller than in the case of a free CDW on the transverse surfaces.

\subsection{Discussion}

We have shown in this article that the macroscopic properties of a CDW can be more surface than bulk dependent especially in thin samples, in particular due to transverse pinning that was either neglected or little considered until now. However, this conclusion raises the question of macroscopic samples depending so much on it surfaces. Why is the bulk CDW phase so closely associated to boundary conditions?

The surfaces of many CDW systems have been studied especially by STM. The CDW phase is present at the top layer and can display a perfect  long range order over hundreds of nanometers, like in the quasi-1D Rb$_{0.3}$MoO$_3$ \cite{brun2005charge}, K$_{0.9}$Mo$_6$O$_{17}$ \cite{mallet1999contrast}, TaS$_3$\cite{gammie1989scanning}, NbSe$_3$\cite{brun2010surface,brun2009scanning,brazovskii2012scanning} and in the quasi-2D TbTe$_3$ \cite{fang2007stm,fu2016multiple} and 1T-TaS$_2$ \cite{burk1991charge}. Moreover, several groups observed a surface CDW using grazing incident X-ray diffraction in NbSe$_2$ \cite{murphy2003surface} and K$_{0.3}$MoO$_3$ \cite{PhysRevB.42.8791}. 

The surface and volume properties of CDW materials may be different but are, however, systems dependent. Although similar CDW properties were reported at the surface and in the bulk of K$_{0.3}$MoO$_3$ by grazing-incidence x-ray diffraction\cite{PhysRevB.42.8791} (the T$^{cdw}_c$ and the order parameter remain identical), this is not the case in other compounds in which the CDW phase in the bulk is different from the surface\cite{ru2008charge} like in TbTe$_3$, on which a second CDW phase along the $a$ axis is measured by STM\cite{fu2016multiple}. In NbSe$_2$, the surface transition was measured at a temperature larger by $1.4\pm0.6$K from the one measured in the bulk. In NbSe$_3$, Brun {\it et al.} \cite{brun2010surface} measured a surface CDW transition temperature ($T_{c}^S = 70-75$K) $15$K above the bulk transition ($T_{c}^{bulk} = 59K$), indicating a different CDW amplitude between surface and bulk. 

Several explanations were given in the literature for the origin of surface pinning in CDW system. Feinberg and Friedel proposed a CDW frontal pinning in the case of rough sample surfaces or, as a second mechanism, a condensation of electrons near the surface if the CDW wavefronts are not perpendicular to the sample transverse surfaces \cite{Feinberg,schlenker1989low}.

The diffraction measurements showing a systematic and identical pinning over so large distances\cite{PhysRevB.101.125122}, however, tend to show that surface pinning is not due to extrinsic surface defects but rather to the intrinsic nature of CDW. As proposed by Yetman and Gill \cite{YETMAN1987201} and in \cite{belle2019}, this effect could be associated to a commensurate CDW pinned to the crystal lattice at the surface in contrast to an incommensurate CDW in the bulk. Gammie et al. measured the surface CDW modulation in TaS$_3$ to be approximately at the commensurate value $4c_0 \times 10b_0$ \cite{PhysRevB.40.11965} where $c_0$ and $b_0$ are the crystal lattice parameters. Brun et al. measured by STM the CDW wavevector in Rb$_{0.3}$MoO$_3$ at the commensurate value $\vec{q}_{cdw}= \pm0.25\vec{b}^*+(\vec{a}+2\vec{c})^*$ \cite{PhysRevB.72.235119}. Finally several STM measurements from the literature indicate a surface CDW wavevector in NbSe$_3$ at $\vec{q}_{cdw}=0.24\vec{b}^*$ \cite{brun2010surface,brazovskii2012scanning,brun2009scanning} close to the commensurate value $0.25\vec{b}^*$. Unfortunately, the error bars of $\vec{q}_{cdw}$ are not indicated, hence, to our knowledge, no  experimental evidence can either support or deny this assumption yet mainly because the surface local probes lack the desired q-resolution.

\section{Conclusion}

In conclusion, we provide here an analytical expression for the CDW phase and threshold field, when the system is submitted to an applied current and constrained by boundary conditions including transverse surface pinning as observed in a recent experiment\cite{PhysRevB.101.125122}.  
The threshold field $E_{th}$ is obtained from the 3D phase  as a function of the sample dimensions and the phase slip process. We show that the CDW deformations appear at larger fields when surface pinning is taken into account, which leads to an increase of threshold fields for small samples. The solution correctly describes the threshold field $E_{th}$ measured in several CDW systems as a function of both, lengths and sample cross sections. 

In addition, this study  shows that bulk impurity pinning, usually introduced in theoretical models as a phenomenological parameter to better fit the data, is not necessary to predict the threshold dependence on the sample dimensions. This threshold field behavior can be explained by the only means of pinning from lateral surfaces without consideration of bulk properties. On the other hand, in the case of thick samples, it seems that bulk impurity pinning still has to be taken into account to predict the temperature dependence of $E_{th}$ \cite{maki1989impurity}.

The CDW dynamics for thin samples is surprisingly mainly controlled by surface effects and less by the volume, despite the comparatively large volumes scales. This strong boundary effect between surfaces tens of micrometers apart can not occur without an extraordinary long range order of the CDW. This effect is consistent with the experiment reported in \cite{PhysRevB.101.125122} that observes a continuous deformation of the CDW wave fronts spanning  a distance of 4 orders of magnitude larger than the CDW wavelength. 

This study also raises the question on the origin of the strong surface pinning effect. As discussed in the text, surface pinning is most probably a intrinsic property of CDW systems and could be related to the loss of incomensurability at the surface. In this framework, the impressive CDW deformation under field would then be a consequence of a coexisting  incommensurate CDW in the volume with commensurate CDW at the surfaces.

The model proposed in this paper highlights the importance of surface effects on sliding properties. The surface state is undoubtedly an important parameter to explain the threshold evolution under mechanical strain \cite{ZYBTSEV201534}. Finally, we show here that surface pinning must be taken into account to correctly describe the threshold field and thus the soliton nucleation at the electrical contact. If transverse pinning plays a significant role for the threshold, it could also be an important parameter to determine the soliton propagation in the sample. Pinning of the soliton could explain the CDW hysteresis effect with current \cite{mihaly1986dielectric,mihaly1984onset,zettl1982onset} as well as the observed decrease of the transverse CDW satellite peak width measured in diffraction \cite{danneau2002motional,belle2019} and a complete theoretical study of this propagation including surface pinning is necessary.

\section{Author contribution statement}

The four authors contributed to the CDW phase solution and the comparison with resistivity experiments from the literature. E.B., V.L.R.J and D.L.B. participated to the experiment presented in Figure \ref{fig:ComparisonKmap}. All four authors contributed to the writing of the paper.

\appendix
\numberwithin{equation}{section}
\section{Description of the phase-slip process}
\label{appendix:phase-slip}

In this section, we describe the phase-slip process defining the electric field threshold Eq.\ref{phaseSlipEq}. As discussed in the introduction, solitons, transporting charges from one contact to the other, are created at the electrical contact when the field exceeds $E_{th}$. However, due to the topology of these $2\pi$ Sine-Gordon phase solitons \cite{peyrard2012physique}, they can not nucleate without destroying the CDW order and the CDW amplitude needs to locally drop to zero. The least energetic way to create a soliton is through a phase vortex ring, also referred to as a CDW dislocation loop \cite{Feinberg}, which appears spontaneously under strain and increases in size until it vanishes on the sample surface, leaving behind it a soliton. This is the so-called phase-slip process that we describe here, starting from the phase vortex configuration.

\begin{figure}
\resizebox{0.47\textwidth}{!}{\includegraphics{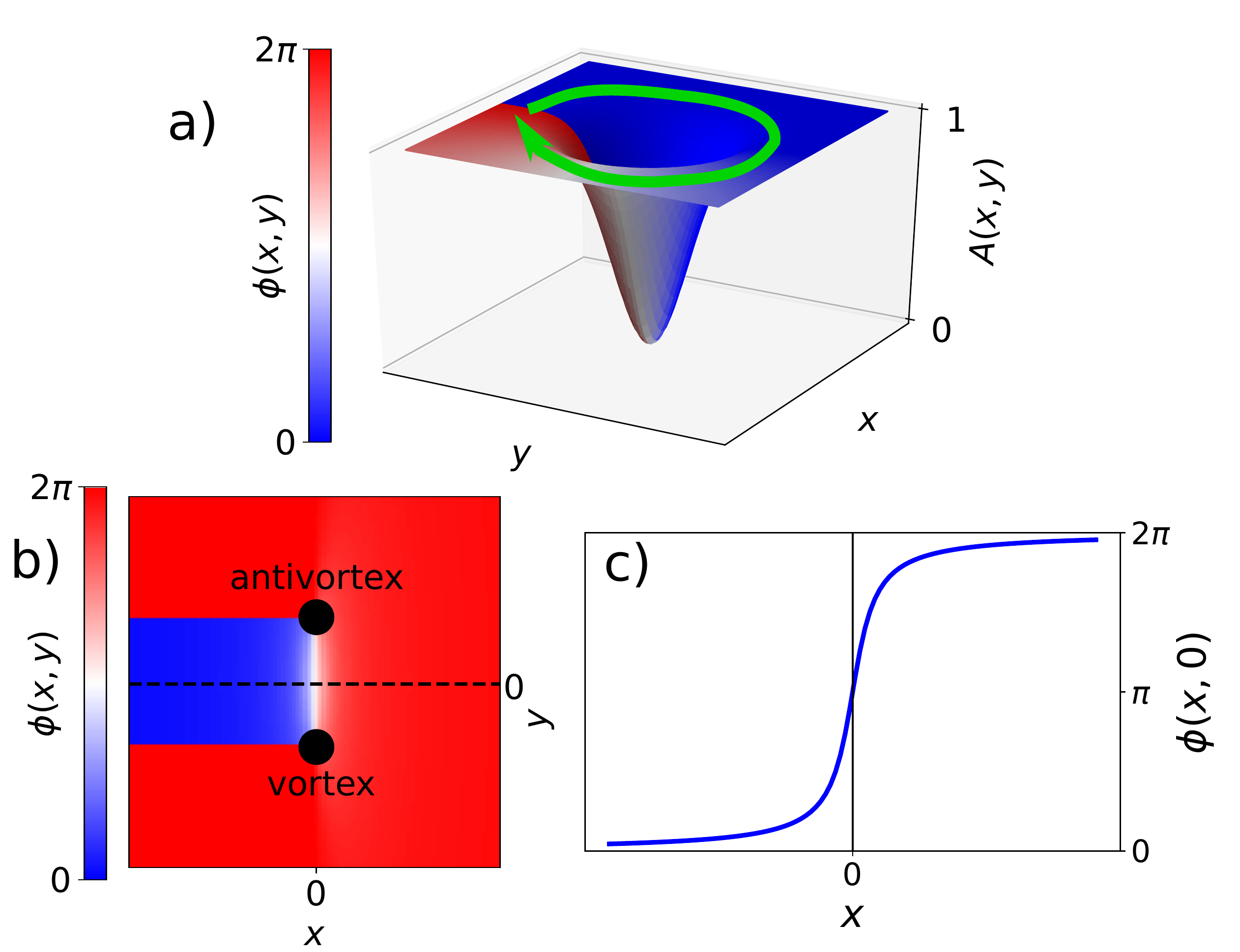}}
\caption{a) 2D topological phase vortex where the CDW amplitude $A$ corresponds to the vertical axis while the CDW phase $\phi$ is displayed in color gradient. Following the green arrow path, $\phi$ increases by $2\pi$. b) Vortex-antivortex pair configuration. c) $\phi$ profile along the black dashed line in c) showing that a topological $2\pi$ soliton is present on each atomic chain inside the vortex-antivortex pair.}
\label{fig:App1}       
\end{figure}

A CDW phase vortex is shown in Fig.\ref{fig:App1}a in which the phase $\phi(\vec{r})$ increases by $2\pi$ when following the green path around the vortex center. In order to avoid an infinite $\phi$ derivative in the center, meaning an infinite elastic energy, the CDW amplitude $A$ drops exponentially to zero at the center as shown in the vertical axis of Fig.\ref{fig:App1}a. Due to its topological property, a vortex configuration can only be created in the CDW volume as a vortex-antivortex pair, where an antivortex is a vortex of opposite chirality. The vortex and antivortex nucleate at the same position and then split in two to create a CDW phase configuration as illustrated in Fig.\ref{fig:App1}b where the vortex centers are schematized as black dots and the $\phi$ value in color. The  $\phi$ profile through the vortex-antivortex configuration contains a 2$\pi$ phase shift (see Fig.\ref{fig:App1}b-c). Hence, every atomic chains located in between the vortex and the antivortex have a CDW soliton. This vortex-antivortex nucleation is the least energetic way to create topological $2\pi$ solitons in the CDW bulk.

\begin{figure}
\resizebox{0.45\textwidth}{!}{\includegraphics{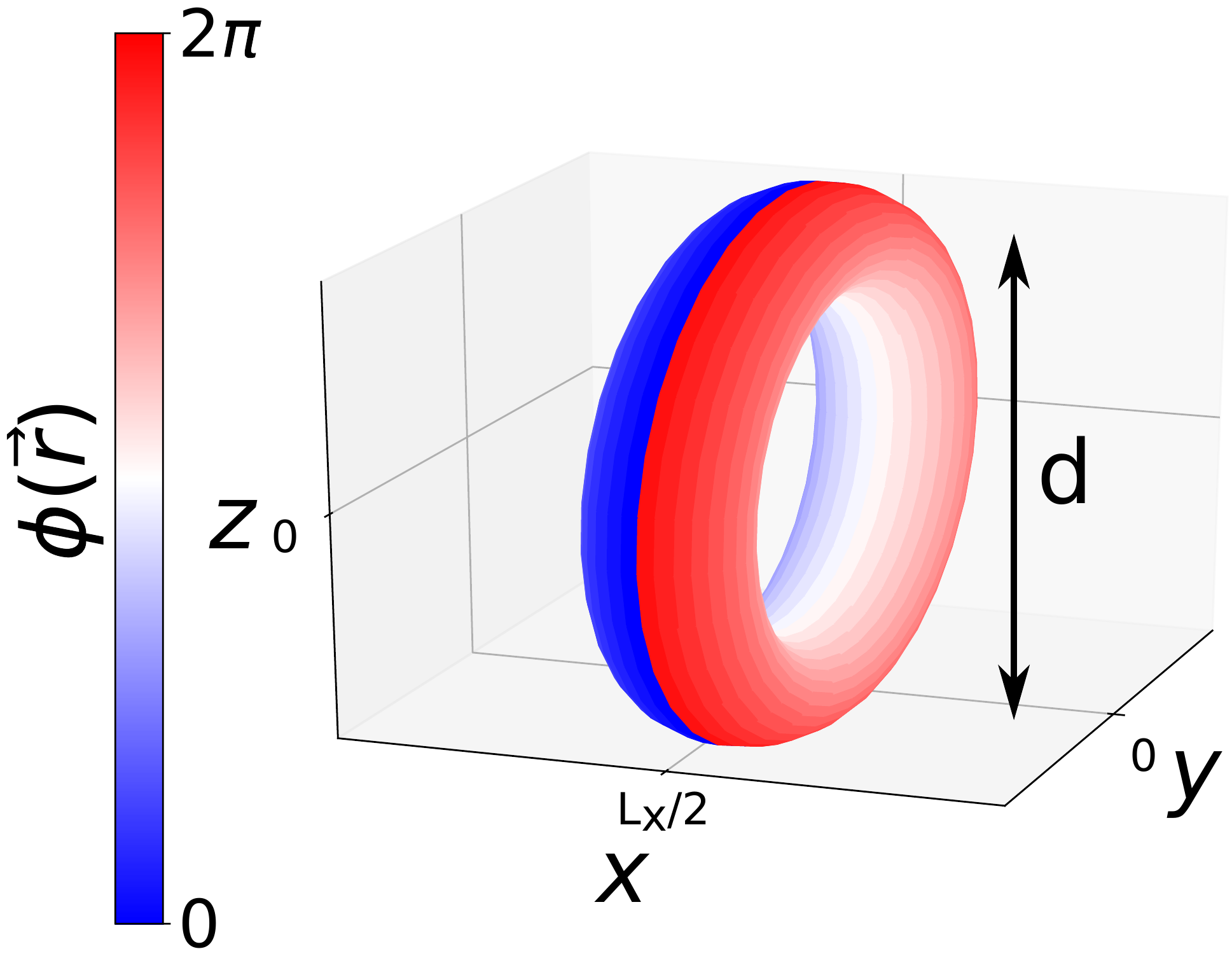}}
\caption{3D vortex ring configuration consisting in a ring of vortex-antivortex pairs. The CDW phase $\phi$ is shown in color gradient on a surface at a fixed distance from the ring perimeter. Every atomic chains inside the ring contains a topological $2\pi$ soliton. The 2D case (Fig.\ref{fig:App1})  corresponds to the section of this 3D ring case in the plane $(x,y,z=0)$.}
\label{fig:App2}       
\end{figure}

The generalization in 3D  is a vortex ring as presented in Fig.\ref{fig:App2} where the CDW phase $\phi$ is shown by a color gradient. This ring corresponds to a continuous vortex-antivortex pairs configuration where the 2D case Fig.\ref{fig:App1}b corresponds to a section of the 3D case in the plane $(x,y,z=0)$. Each atomic chain inside this ring contains a CDW soliton as in the 2D case (Fig.\ref{fig:App1}b and c).

\begin{figure}
\resizebox{0.43\textwidth}{!}{\includegraphics{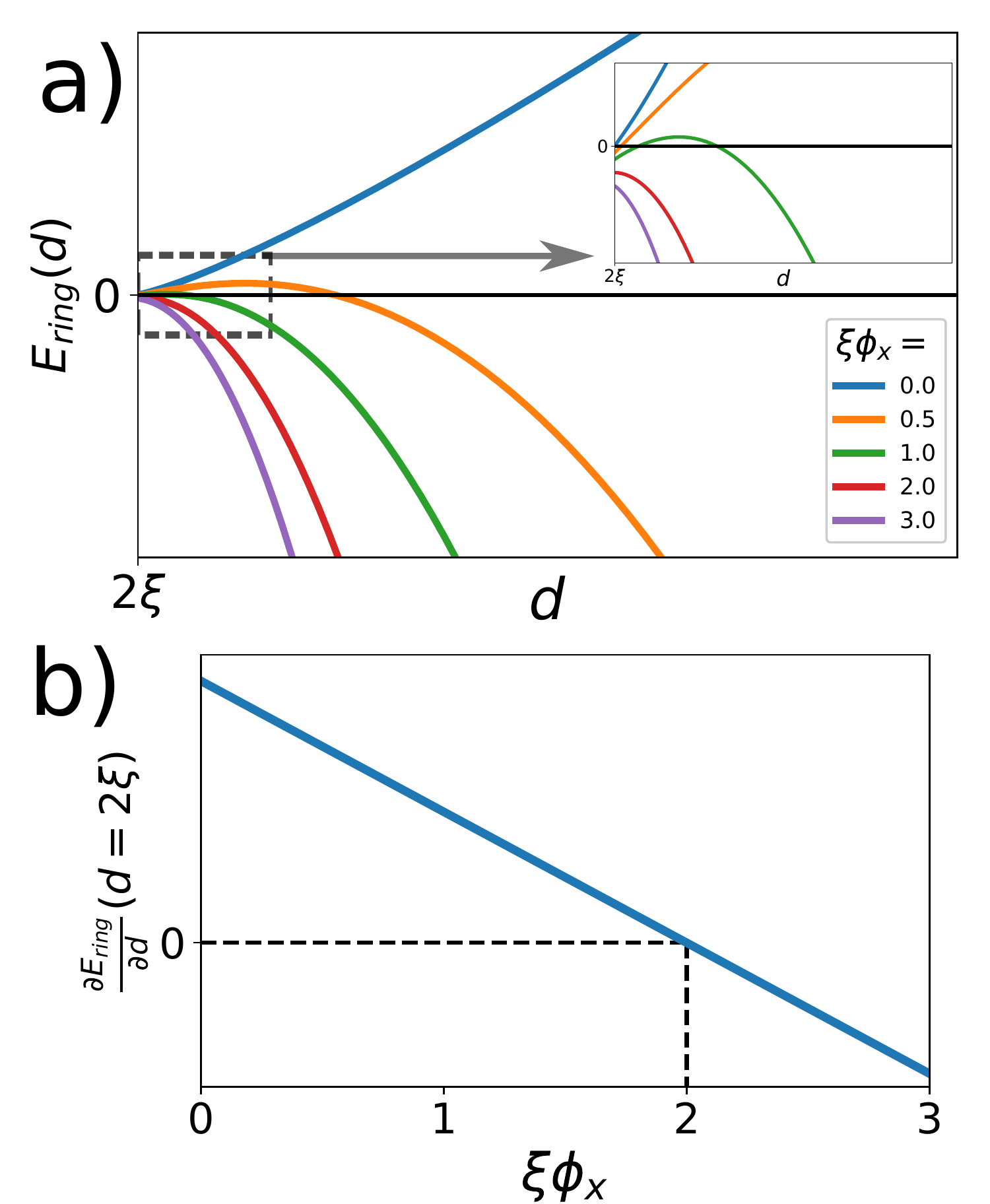}}
\caption{a)Energy of a vortex ring in the presence of a strain $\propto\phi_x$ induced by the applied electric field. inset: zoom on the region of small $d$. b) derivative of the ring energy at the smallest ring size $d=2\xi$. For $\xi\phi_x<2$, $\frac{\partial E_{ring}}{\partial d}(d=2\xi)$ is positive, thus an energy barrier prevents the ring size $d$ to increase. However, for $\xi\phi_x>2$, $\frac{\partial E_{ring}}{\partial d}$ is negative $\forall d$, hence the barrier disappear and $d$ grows spontaneously until the ring perimeters reaches the sample surfaces and vanishes, leaving behind a topological $2\pi$ soliton on each atomic chain at the electrical contact. This phenomenon is the "phase-slip process".}
\label{fig:App3}       
\end{figure}

The vortex ring induces a phase gradient leading to the elastic energy\cite{belle2019}:
\begin{equation}
E_{\text{ring}}(d) \propto 4\pi^2d\ln\left(\frac{d-\xi}{\xi}\right)-\pi^2d^2\phi_x
\end{equation}
where $d$ is the vortex diameter and $\xi$ the CDW coherence length. As the vortex center radius is $\xi$, the ring energy isn't defined for a diameter $d\leq 2\xi$.  The elastic energy $E_{\text{ring}}$ without strain ($\phi_x$=0) is shown as function of the diameter $d$  in blue in Fig.\ref{fig:App3}a. We observe that $E_{\text{ring}}$ is always positive and increases for large $d$, thus the vortex ring can not nucleate spontaneously in a CDW at equilibrium. However, this is no more the case under an external electric field inducing a CDW longitudinal strain $\phi_x\neq 0$ (seed Fig.\ref{fig:3}b). The ring energy decreases in the presence of a non-zero $\phi_x$ as shown in Fig.\ref{fig:App3}a and for large ring size ($d\gg\xi$) $E_{\text{ring}}$ is negative and the ring nucleation becomes favorable.

However if $\xi\phi_x<2$ (blue, orange and green curves in Fig.\ref{fig:App3}a) the energy derivative at smallest ring diameter $\frac{\partial E_{ring}}{\partial d}(d=2\xi)$ is positive, hence an energy barrier forbids the ring to increase to larger diameters and to nucleate. Yet, as $\phi_x$ increases, $\frac{\partial E_{ring}}{\partial d}(d=2\xi)$ decreases linearly as shown in Fig.\ref{fig:App3}b until it becomes negative for $\xi\phi_x>2$, hence the energy barrier disappear and the ring nucleates spontaneously. As shown in the violet curve of Fig.\ref{fig:App3}b, for $\phi_x>2/\xi$, $E_{ring}$ is always decreasing, therefore the diameter $d$ increases until the ring reaches the sample borders and vanishes, leaving behind it a charged soliton on each atomic chain at the electrical contact. This is the phase-slip process.

We see from this argument that the ring nucleates where the longitudinal derivative $\phi_x$ is maximum, i.e. at the electrical contact and as far as possible from the lateral surface in $(x=L_x/2,y=0,z=0)$with our model (see Fig.\ref{fig:3}b) and Fig.\ref{fig:App4}b. The soliton nucleation with the appearance of the non-linear CDW current occurs when the applied electric field is such that the phase longitudinal derivative is larger than a critical value $ \phi_x(\frac{L_x}{2},0,0)> \phi'_c \equiv 2/\xi$, hence defining the threshold field Eq.\ref{phaseSlipEq}. 

Finally, we considered here the strain to be constant locally, that is $\phi_ x$ is constant in the calculation of $E_{\text{ring}}(d)$ shown in Fig.\ref{fig:App3}. This approximation is correct since the typical length of the derivative variation is of the order of the sample dimensions $L_x,L_y,L_z$ as observed in Fig.\ref{fig:3}b). Since $E_{\text{ring}}$ reaches a large negative value for $d>>\xi$ and $\phi_x(\frac{L_x}{2},y,z)\geq 0\, (\forall \, y,z)$, the spatial variation of the CDW strain won't prevent the ring size to spontaneously increase.

\section{Uniqueness theorem}
\label{appendix:uniqueness}

The image charges method is often used to solve electrostatic problems with boundary conditions by adding charges  strategically placed to enable the Laplace equation to be more easily solved. This method is valid in electrostatic since the Poisson equation satisfies the uniqueness theorem regarding the solutions gradient (electric field). The use of the image charge method in our case with contacts and surface pinning is justified if the inhomogeneous screened Poisson equation Eq.\ref{equationChangeVariable} also satisfies this theorem.

Let's consider two different solutions $\phi_1(\vec{r})$ and $\phi_2(\vec{r})$ of the Eq.\ref{equationChangeVariable} both satisfying the boundary conditions Eq.\ref{conditionsChangeVariable} and the difference $\psi=\phi_1-\phi_2$ that must be zero if the solution is unique.
This phase difference $\psi$ fulfill the homogeneous screened Poisson equation along side with the Dirichlet conditions Eq\ref{conditionsChangeVariable}: 
\begin{equation}\label{uniqueness1}
(\Delta-\omega^2) \psi =0
\end{equation}
The only solution to this equation is the trivial solution $\psi = 0$ since in this case the Laplacian is negative definite while $\omega^2 > 0$\cite{Laplacian}. In the following, we show a step by step demonstration of this trivial solution. Let's consider the relations:
\begin{align}
\vec{\nabla}\left(\psi\vec{\nabla}\psi\right)&= \left(\vec{\nabla}\psi\right)^2+\psi\Delta\psi  \nonumber \\
&= \left(\vec{\nabla}\psi\right)^2+\omega^2\psi^2 \label{uniqueness2}
\end{align}
where Eq.\ref{uniqueness1} is used in the second line. By integrating Eq.\ref{uniqueness2} over the CDW sample volume $\mathcal{V}$ and using the divergence theorem one finds:
\begin{align}
\int_\mathcal{V} \vec{\nabla}\left(\psi\vec{\nabla}\psi\right) &= \int_\mathcal{V} \left[\left(\vec{\nabla}\psi\right)^2+\omega^2\psi^2 \right] d^3\vec{r} \nonumber \\
\int_{\partial \mathcal{V}} \psi\vec{\nabla}\psi &= \int_\mathcal{V} \left[\left(\vec{\nabla}\psi\right)^2+\omega^2\psi^2 \right] d^3\vec{r} \label{uniqueness3}
\end{align}
where the surface $\partial \mathcal{V}$ corresponds to the CDW boundaries. Since $\psi$ satisfies Eq.\ref{conditionsChangeVariable}, $\psi=0$ on the boundaries and the left hand side of Eq.\ref{uniqueness3} is zero. Therefore, each squares terms on the right hand side must be zero to vanish the integral. Hence  $\psi(\vec{r})=0 \, (\forall \, \vec{r)}$, proving the unicity of the solution.

\section{Step-by-step construction of the image charge and CDW phase solution in 1D and 2D}
\label{appendix:stepImageCharge}

\begin{figure}
\resizebox{0.48\textwidth}{!}{\includegraphics{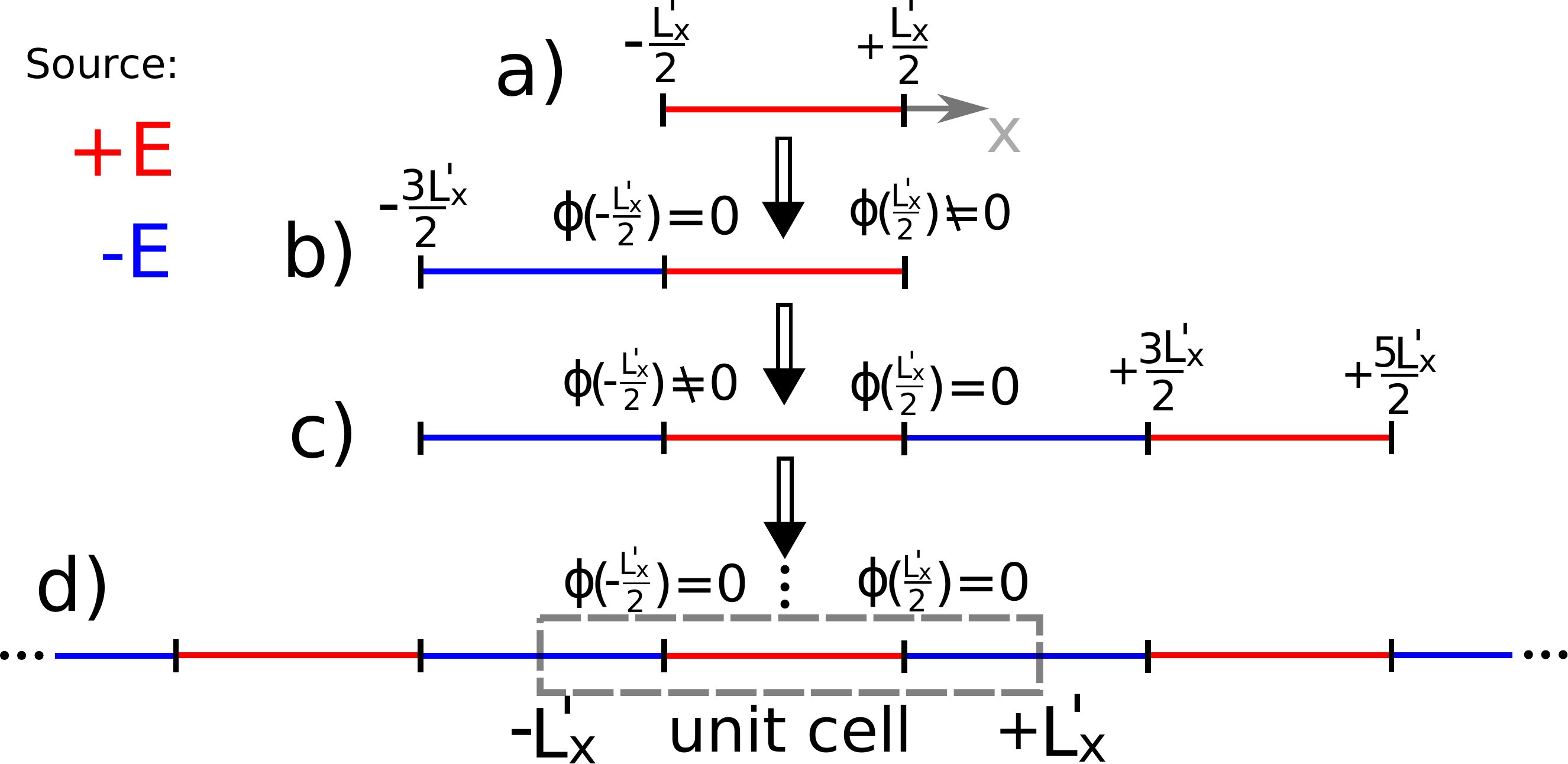}}
\caption{The uniform charge density $+E$ in the CDW system used in the image charge method (see Eq.\ref{equationChangeVariable}) is displayed as a red line for $x'\in[-\frac{L_x'}{2},\frac{L'}{2}]$ while the negative image charges are displayed in blue. b), c) and d) present the step-by-step construction of the image charge density array described in this appendix.}
\label{fig:AppImageCharge}       
\end{figure}

A step-by-step construction of the image charge density, for the 1D case, satisfying $\phi(\pm \frac{L_x'}{2})=0$ is shown in Fig.\ref{fig:AppImageCharge}. The first step is to consider a uniform source term $+E$ in the bulk of the sample (the red line in Fig.\ref{fig:AppImageCharge}a where the source density is $\rho_a(x') = E\times\Pi\left(\frac{x'}{L_x'}\right)$  where $\Pi$ is the gate function defined $\Pi(x) = \Theta(\frac{1}{2} - \lvert x \rvert)$ with $\Theta$ being the Heavyside function.

The second step is to consider an "anti-mirror" located at $x'=-\frac{L'}{2}$  to create an artificial negative image charge density $-E$ located in the region $-\frac{3L_x'}{2}<x'<-\frac{L'}{2}$ that fulfills the boundary condition $\phi(-\frac{L_x'}{2})=0$ as shown in Fig.\ref{fig:AppImageCharge}b. At the end of this step, the total charge density is $\rho_b(x') = \rho_a(x')-E~ \Pi\left(\frac{x'+L_x'}{L_x'}\right)$

In order to satisfy the second boundary condition $\phi(\frac{L'}{2})=0$,  the procedure has to be repeated by adding two additional image charge densities, $-E$ at $\frac{L_x'}{2}<x'<\frac{3L_x'}{2}$ and $+E$ at $\frac{3L_x'}{2}<x'<\frac{5L_x'}{2}$ as shown in Fig.\ref{fig:AppImageCharge}c. However, while this new charge density impose $\phi(\frac{L_x'}{2})=0$, the first boundary condition is lost since $\phi(-\frac{L_x'}{2})\neq 0$  (see Fig.\ref{fig:AppImageCharge}c). Therefore, we must repeat the process back and forth between the left and right side of the sample  until reaching an infinite periodic total charge density $\rho(x')$ shown in Fig.\ref{fig:AppImageCharge}d which can expressed by the infinite sum:

\begin{equation}
\rho(x') = \sum\limits_{p=-\infty}^{+\infty} \rho_{unit}(x'-p2L_x') 
\end{equation}
where the density of the unit cell $\rho_{unit}$ (see the grey rectangle in Fig.\ref{fig:AppImageCharge}d) reads:
\begin{equation}\label{rhoUnit1D}
    \begin{split}
    \rho_{unit}(x') = E\left[-\Pi\left(\frac{2\left(x'+\frac{3L_x'}{4}\right)}{L_x'}\right)+\Pi\left(\frac{x'}{L_x'}\right) \right. \\
    \left. -\Pi\left(\frac{2\left(x'-\frac{3L_x'}{4}\right)}{L_x'}\right)\right]
    \end{split} 
\end{equation}

The phase is the space convolution $ \phi(x')=[G\star \rho](x')$ which is easier to express in Fourier space as a product $\phi(q)=G(q)\times \rho(q)$. The Fourier transform of the unit cell density is:
\begin{equation}\label{rhoUnitFourier1D}
\rho_{unit}(q) = E\frac{8}{q}\sin^2\left(\frac{qL_x'}{4}\right)\sin\left(\frac{qL_x'}{2}\right)
\end{equation}
and since the total charge density $\rho(x')$ is an infinite periodic array, its Fourier transform is an infinite sum:
\begin{equation}
\rho(q)= \rho_{unit}(q)\frac{\pi}{L_x'}\sum\limits_{h=-\infty}^{+\infty}\delta\left(q-h\frac{\pi}{L_x'}\right)
\end{equation}

Finally, the phase solution is the inverse Fourier transform of $G(q)\rho(q)$. One finds an infinite sum that can be simplified using first $\rho_{unit}(-h\frac{\pi}{L_x'}) = \rho_{unit}(+h\frac{\pi}{L_x'})$ and  then $\rho(2n\frac{\pi}{L_x'})=0$ and $\rho((2n+1)\frac{\pi}{L_x'})=\frac{4EL'(-1)^n}{\pi(2n+1)}$. We finally obtain the following expression for the CDW phase in 1D:

\begin{equation}
\phi(x') = -\frac{4E}{\pi}\sum\limits_{n=0}^{+\infty}\frac{(-1)^n\cos\left[(2n+1)\pi\frac{x'}{L_x'}\right]}{(2n+1)\left\lbrace \left[ (2n+1)\frac{\pi}{L_x'}\right]^2+\omega^2 \right\rbrace}
\end{equation}

\begin{figure}
\resizebox{0.48\textwidth}{!}{\includegraphics{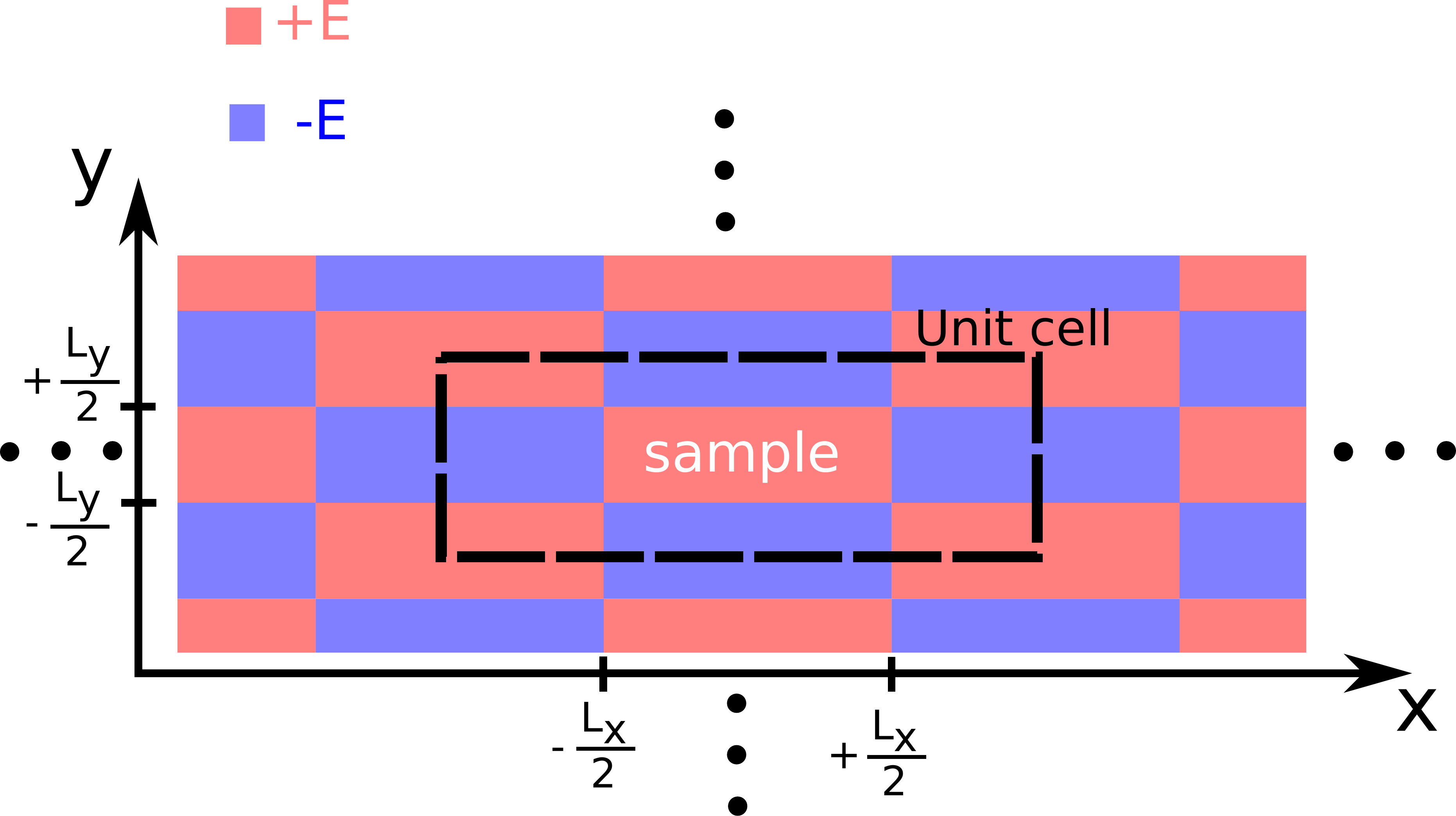}}
\caption{2D periodic array of image charge density with the unit cell represented by the black dashed-line rectangle. Red (blue) rectangles correspond to a uniform positive (negative) charge density $+E$ ($-E$).}
\label{fig:AppImageCharge2D}       
\end{figure}

The construction of the image charge density in 2D is similar to the procedure used in 1D. The CDW sample of size $L'_x\times L'_y$ has a uniform charge density $+E$ and an infinite array of periodic image charges is built to fulfill the boundary conditions $\phi(\pm\frac{L'_x}{2},y')=\phi(x',\pm\frac{L'_y}{2})=0$ (see Fig.\ref{fig:AppImageCharge2D}). The Fourier transform of the  unit cell of the two-dimensional periodic density is:

\begin{equation}\label{rhoUnitFourier2D}
\begin{split}
\rho_{unit}(q_x,q_y) = E\frac{64}{q_xq_y}\sin\left( \frac{q_xL'_x}{2}\right)\sin^2\left( \frac{q_xL'_x}{4}\right) \times \\
\sin\left( \frac{q_yL'_y}{2}\right)\sin^2\left( \frac{q_yL'_y}{4}\right)
\end{split}
\end{equation}
The Fourier transform of the infinite electronic crystal becomes a double sum of Dirac products and after simplification and using the same symmetry arguments as in 1D between positive and negative terms and the fact that $\rho$ is zero for even terms of the sum, the 2D CDW phase solution can be written as: 
\begin{equation}
\begin{split}
\phi(x',y') = -\frac{16E}{\pi^2}\sum\limits_{n_{x,y}=0}^{+\infty}\frac{(-1)^{n_x+n_y}}{(2n_x+1)(2n_y+1)} \times  \\
\frac{\cos\left[(2n_x+1)\pi\frac{x'}{L'_x}\right]\cos\left[(2n_y+1)\pi\frac{y'}{L'_y}\right] }{ \left[(2n_x+1)\frac{\pi}{L'_x}\right]^2+\left[(2n_y+1)\frac{\pi}{L'_y}\right]^2+\omega^2}
\end{split}
\end{equation}

\section{The CDW phase deformation in 3D}
\label{appendix:3Dsolution}

Following the same procedure as in 1D (Eq.\ref{1DSolutionSum}) and in 2D (Eq.\ref{2DSolutionSum}), the solution of the CDW phase submitted to an electric field with Dirichlet conditions in 3D reads:

\begin{equation}
\begin{split}
\phi(\vec{r}') = -\frac{64E}{\pi^3}\sum\limits_{n_{x,y,z}=0}^{+\infty}\frac{(-1)^{n_x+n_y+n_z}}{(2n_x+1)(2n_y+1)(2n_z+1)} \times  \\
 \frac{\cos\left[(2n_x+1)\pi\frac{x'}{L_x'}\right]\cos\left[(2n_y+1)\pi\frac{y'}{L_y'}\right]\cos\left[(2n_z+1)\pi\frac{z'}{L_z'}\right]}{ \left[(2n_x+1)\frac{\pi}{L'_x}\right]^2+\left[(2n_y+1)\frac{\pi}{L'_y}\right]^2+\left[(2n_z+1)\frac{\pi}{L'_z}\right]^2+\omega^2}
 \label{3DSolutionSum}
\end{split}
\end{equation}

\begin{figure}
\resizebox{0.48\textwidth}{!}{\includegraphics{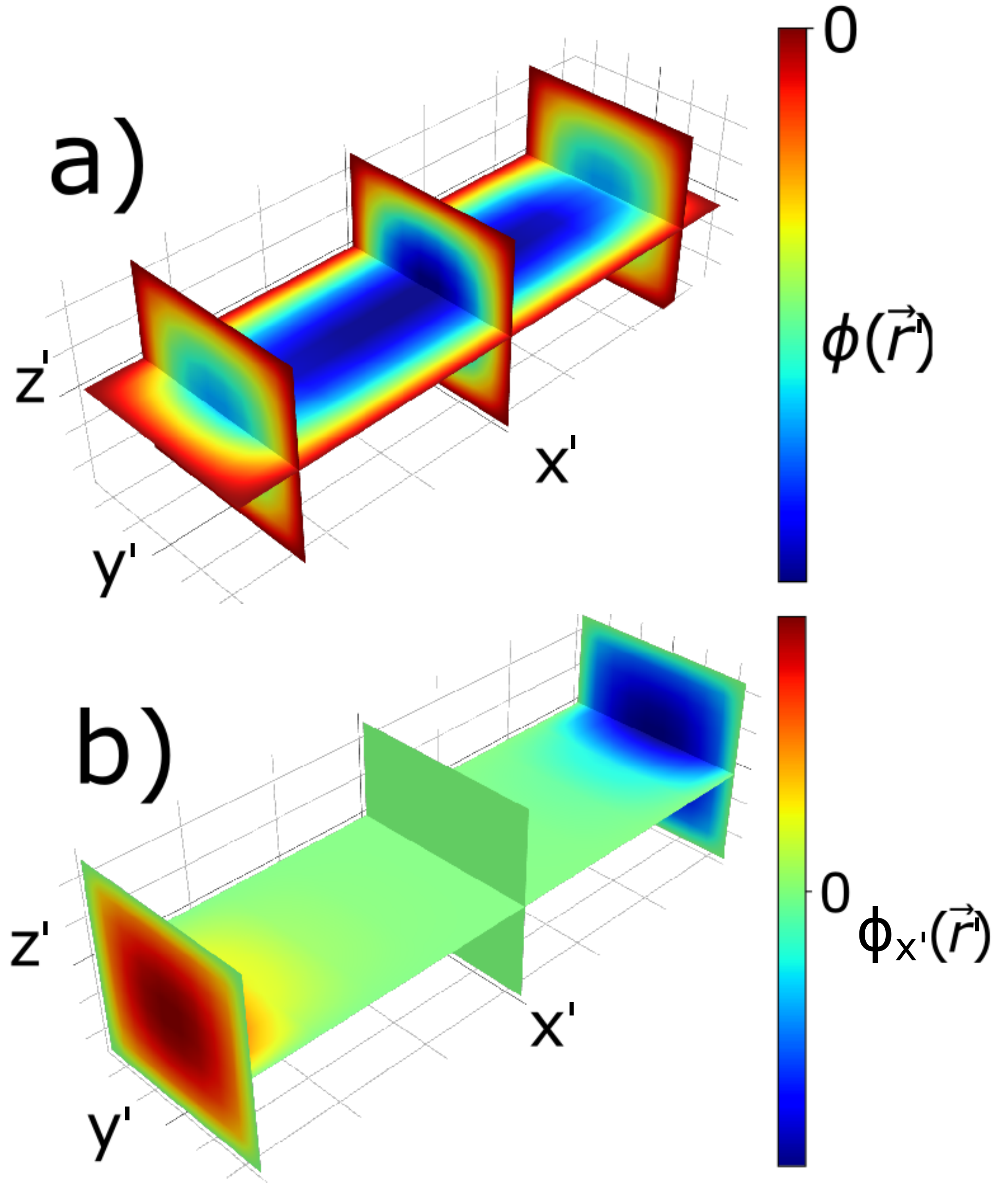}}
\caption{a) The 3D phase $\phi(\vec r)$ solution of the Eq.\ref{equationChangeVariable} with Dirichlet conditions with $\omega = 0$, $E=1$, $L'_x=3$, $L'_y=L'_z=1$ and by summing the first hundred  terms of the sum ($n_x,n_y,n_z\leq 100$). b) The longitudinal derivative $\phi_{x'}$ is displayed in color scale with slices through the sample volume. The strain reaches a maximum at the electrical contact $\vec r=(L_x/2, 0,0)$ where the phase slip occurs and a minimum at the opposite side $\vec r=(-L_x/2, 0,0)$. The extrema are reversed if the current direction is changed.}
\label{fig:App4}       
\end{figure}

The corresponding phase profile $\phi(\vec{r}')$ is shown in Fig.\ref{fig:App4} a) in the 3D volume of the sample. The phase is zero at the sample boundaries as expected while $\phi$ varies mostly in the middle of the sample. The longitudinal derivative $\phi_{x'}$ is close to zero in the middle of the sample and reaches its maximum at the electrical contact where the phase slip occurs (see Fig.\ref{fig:App4}b).  One can also observe that $\phi_{x'}$ drops to zero at the transverse surfaces along y and z. Therefore, as considered in the main text, the phase slip occurs where $\phi_{x'}$ is maximum, at $(x',y',z')=(L_x'/2,0,0)$.

\section{Fitting procedure for the 3D threshold field $E_{th}$ versus the length $L_x$}
\label{fitting}

\begin{figure}
\resizebox{0.4\textwidth}{!}{\includegraphics{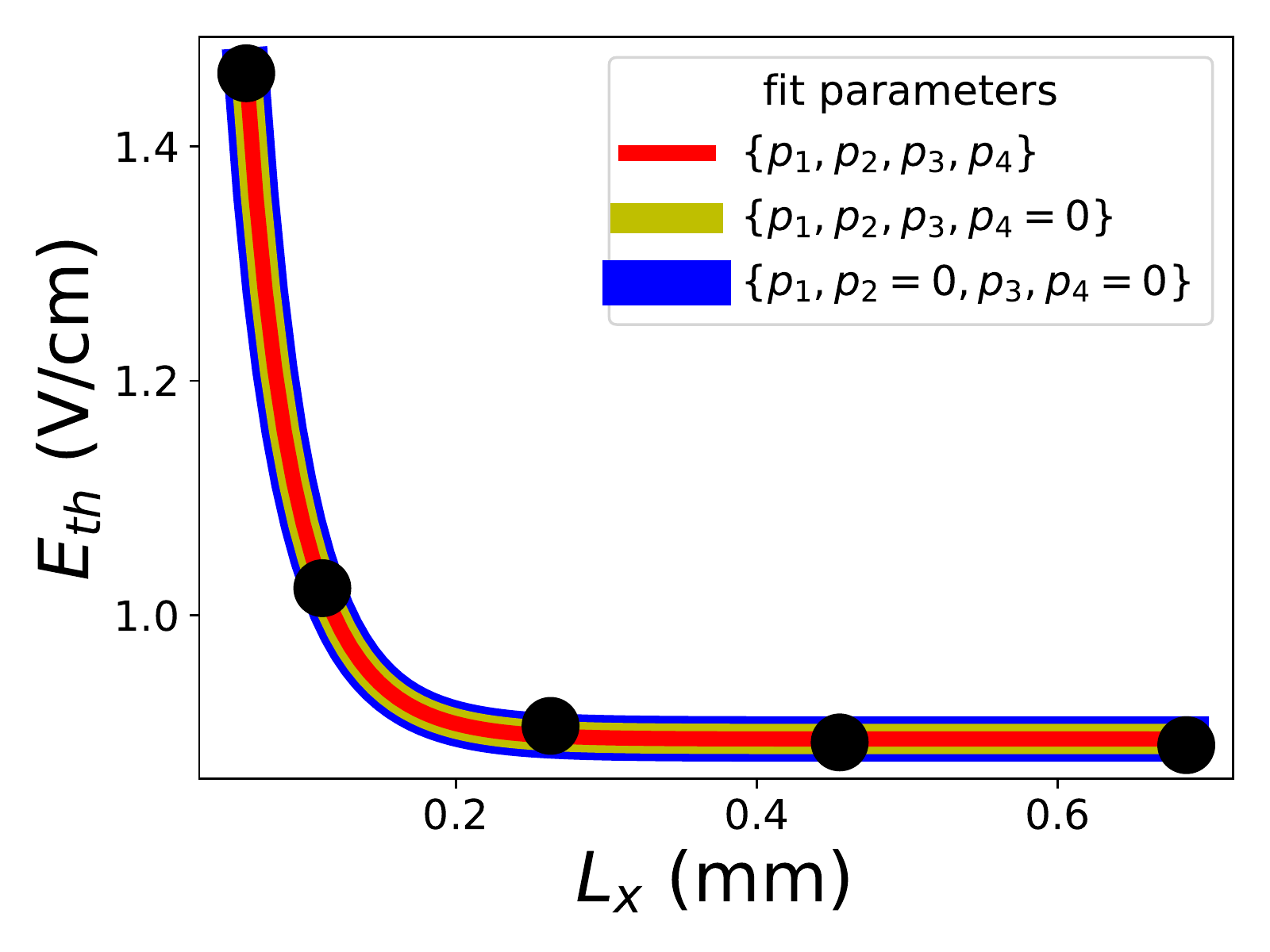}}
\caption{Resistivity data from \cite{PhysRevB.32.2621} fitted by using three different sets of parameters in Eq.\ref{aFitLx}: $\left\lbrace  p_1,p_2,p_3,p_4 \right \rbrace$ (red curve), without bulk pinning $\left\lbrace  p_1,p_2,p_3,0 \right \rbrace$ (yellow curve) and with equal transverse elastic constants ($c_y=c_z$) and thinner than large samples (L$_y\gg$ L$_z$) with $\left\lbrace  p_1,0,p_3,0 \right \rbrace$. The  last fit gives $p_1=0.080 \pm 0.002$ and $p_3=5.5 \pm 0.1$.}
\label{fig:4}       
\end{figure}

In this section, we describe the procedure used to  fit the expression of the threshold field given in  Eq.\ref{Eth} from the experimental resistivity measurements \cite{PhysRevB.32.2621,PhysRevB.29.755,MIHALY1983203}. As discussed in the main text, 4 free parameters are used 
$$\left\lbrace p_1,p_2,p_3,p_4\right\rbrace = \left\lbrace\frac{\phi'_c\pi^3c_x^2}{8\eta}, \frac{c_y}{L_yc_x},\frac{c_z}{L_zc_x},\frac{\omega_0}{\sqrt{2}\pi c_x}\right\rbrace$$
with:
\begin{equation}\label{EthFitLx}
\begin{split}
E^{fit \, L_x}_{th}\left(L_x,\{p_1,p_2,p_3,p_4\}\right) = \frac{p_1}{ L_x} \times \\
\frac{1}{\sum\limits_{n_{y,z}=0}^{+\infty} \frac{(-1)^{n_y+n_z}}{(2n_y+1)(2n_z+1)a^{fit}_{n_y,n_z}} \text{tanh}\left(\frac{\pi a^{fit}_{n_y,n_z}}{2}\right)} 
\end{split}
\end{equation}
with
\begin{equation} \label{aFitLx}
\left(a^{fit \, L_x}_{ny,n_z}\right)^2=\left[(2n_y+1)p_2L_x\right]^2+\left[(2n_z+1)p_3L_x\right]^2+\left(p_4 L_x\right)^2
\end{equation}
We fit  $E_{th}$ as a function of $L_x$ from Prester {\it et al} \cite{PhysRevB.32.2621} for different choices of parameters. 
As a first step, we fit the data by using the 4 parameters $\left\lbrace p_1,p_2,p_3,p_4\right\rbrace$ (see the red curve in Fig.\ref{fig:4}). The data are correctly fitted but the covariance matrix has large diagonal as well as non-diagonal elements showing correlation between parameters, especially between $p_2$ and $p_3$ (see Eq.\ref{aFitLx}).

As discussed in the main text, the characteristics of $E_{th}$ can be reproduced without the need of $\omega_0$. Without bulk pinning $(\omega_0=p_4=0)$, the global $E_{th}$ profile with $L_x$ is also well reproduced (see the yellow curve in Fig.\ref{fig:4}).

In addition, the two transverse elastic constant $c_y$ and $c_z$ are assumed to be the same, as in K$_{0.3}$MoO$_3$ \cite{PhysRevB.43.8421}. Furthermore, the standard NbSe$_3$ dimensions allow us to set  $L_y\gg L_z$ leading to $p_2\ll p_3$. By using only two free parameters $\left\lbrace p_1,p_2=0,p_3,p_4=0 \right\rbrace$, our model can still correctly reproduce the overall behavior of the experimental E$_{th}$ (see the blue line in Fig.\ref{fig:4}). 

\section{Convergence of the $E_{th}$ expression}
\label{appendix:numericalConvergence}

The double sum used in the expression of $E_{th}$ reads (where we set $\omega_0=0$ as in the fits of Fig.\ref{fig:5} and \ref{fig:6})
\begin{equation} 
\label{sumConvergence}
S(k_1,k_2)= \sum\limits_{n_{y,z}=0}^{\infty}\frac{(-1)^{n_y+n_z}\text{tanh}\left( \frac{\pi}{2}a_{n_y,n_z} \right) }{(2n_y+1)(2n_z+1)a_{n_y,n_z}},
\end{equation} 
with
\begin{equation}
a^2_{n_y,n_z}= \left[(2n_y+1)k_1\right]^2+\left[(2n_z+1)k_2 \right]^2
\end{equation} 
and $k_1= \frac{L_x}{c_x}\frac{c_y}{L_y}$ and $k_2= \frac{L_x}{c_x}\frac{c_z}{L_z}$. We evaluate  here the error made on $S$ when a finite number of terms is used in the sum.

\begin{figure}
\resizebox{0.48\textwidth}{!}{\includegraphics{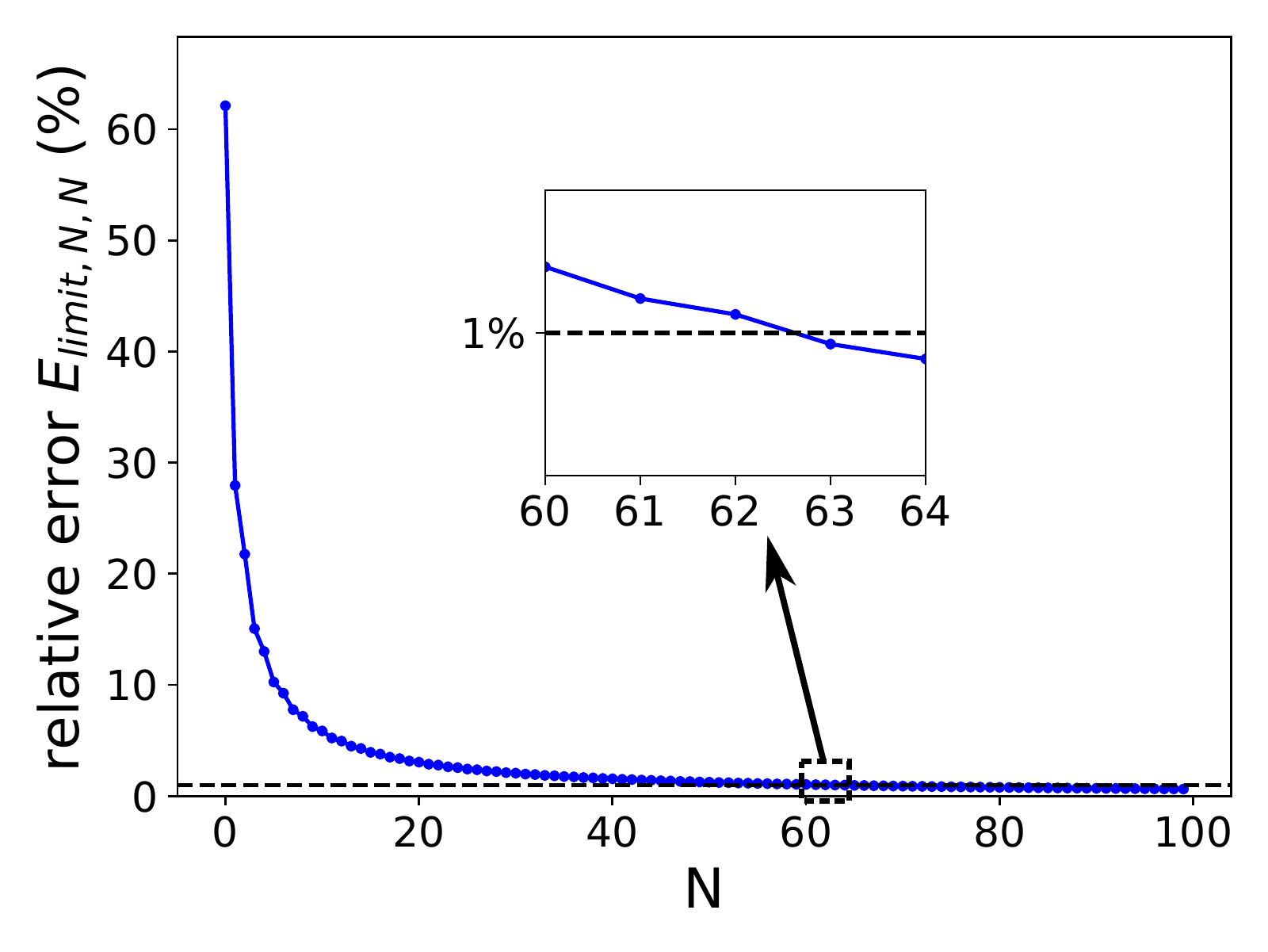}}
\caption{Relative error $E_{\text{limit},N,N}$ (Eq.\ref{limit3DRelativeError}) made in the numerical approximation of Eq.\ref{sumConvergence} in the limit case $k_1=k_2 =0$ and taking into account the first terms for which $n_y,n_z\leq N$, both equal to $N_y=N_z=N$. The contribution of the first $N=63$ terms is necessary to have less than $1\%$ relative error.}
\label{fig:App5}       
\end{figure}

For large $k_1$ and $k_2$ values, the hyperbolic tangent term tends towards one and the double series converges as an alternating sign inverse square $\sim \frac{(-1)^n}{(2n+1)^2}$. However, in the other limit ($k_1,k_2 \ll 1$), the first order expansion of tanh gives: 
\begin{equation} 
\lim\limits_{k_1,k_2 \ll 1}S(k_1,k_2)=\frac{\pi}{2} \sum\limits_{n_{y,z} = 0}^{\infty}\frac{(-1)^{n_y+n_z}}{(2n_y+1)(2n_z+1)},
\end{equation}
that converges much slower as an alternating sign inverse linear convergence. In the following, we will study the sum convergence in this specific case knowing that $S(k_1,k_2)$ converges much faster for the other cases.  The relative error considering the first $N_y$ and $N_z$ terms in the double sum 
\begin{equation}\label{Slimit}
S_{\text{limit},N_y,N_z} = \sum\limits_{\substack{0\leq n_y\leq N_y \\ 0\leq n_z \leq N_z}}\frac{(-1)^{n_y+n_z}}{(2n_y+1)(2n_z+1)}
\end{equation}
can be written as:
\begin{align}
E_{\text{limit},N_y,N_z} \equiv \left|\frac{S_{\text{limit},\infty,\infty}-S_{\text{limit},N_y,N_z}}{S_{\text{limit},\infty,\infty}}\right| \notag \\
= \left|1-\left[1+\frac{2}{\pi}(-1)^{N_y}\Phi\left(-1,1,\frac{3}{2}+N_y\right) \right] \right. \notag  \\
\times \left. \left[1+\frac{2}{\pi}(-1)^{N_z}\Phi\left(-1,1,\frac{3}{2}+N_z\right) \right]\right| \label{limit3DRelativeError}
\end{align}
where $\Phi(z,s,a)$ is the Lerch transcendent function\cite{Lerch,Erdelyi}. This relative error $E_{\text{limit},N,N}$ for $N_y=N_z=N$ is shown in Fig.\ref{fig:App5}.
The sum up to the $63^{th}$ term both in $n_y$ and $n_z$ is necessary to obtain an error less than $1\%$. This constraint has been applied in all fits of the main text (Fig.\ref{fig:5} and \ref{fig:6}). 

\begin{figure}
\resizebox{0.45\textwidth}{!}{\includegraphics{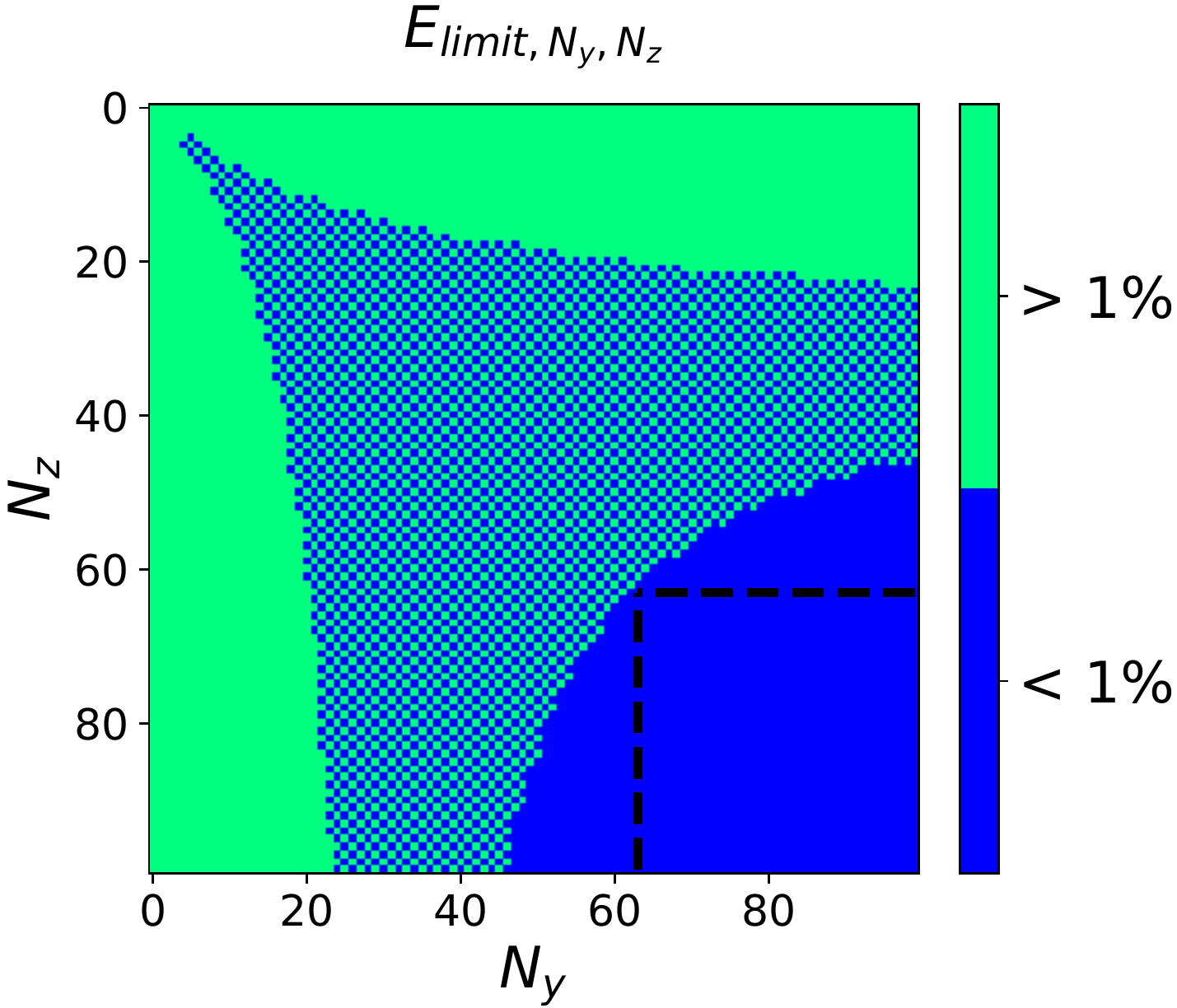}}
\caption{Numerical relative error Eq.\ref{limit3DRelativeError} in the limit case $k_1=k_2=0$ summing terms $n_y$ ($n_z$) from $0$ to $N_y$ ($N_z$). The blue region ($N_y,N_z\geq 63$) corresponds to relative errors less than $1\%$. The oscillatory small blue regions where $N_y,N_z< 63$ are due to the alternating sign in Eq.\ref{sumConvergence}.}
\label{fig:App6}       
\end{figure}

The relative error $E_{\text{limit},N_x,N_z}$ for $N_y \neq N_z$ is shown in Fig.\ref{fig:App6} where the blue region corresponds to relative errors less than $1\%$. As in Fig.\ref{fig:App5}, $E_{\text{limit},N_y,N_z}< 1\%$ when $N_y$ and $N_z$ are both larger than $63$ (region in the lower right part of Fig.\ref{fig:App6}).

\begin{figure}
\resizebox{0.48\textwidth}{!}{\includegraphics{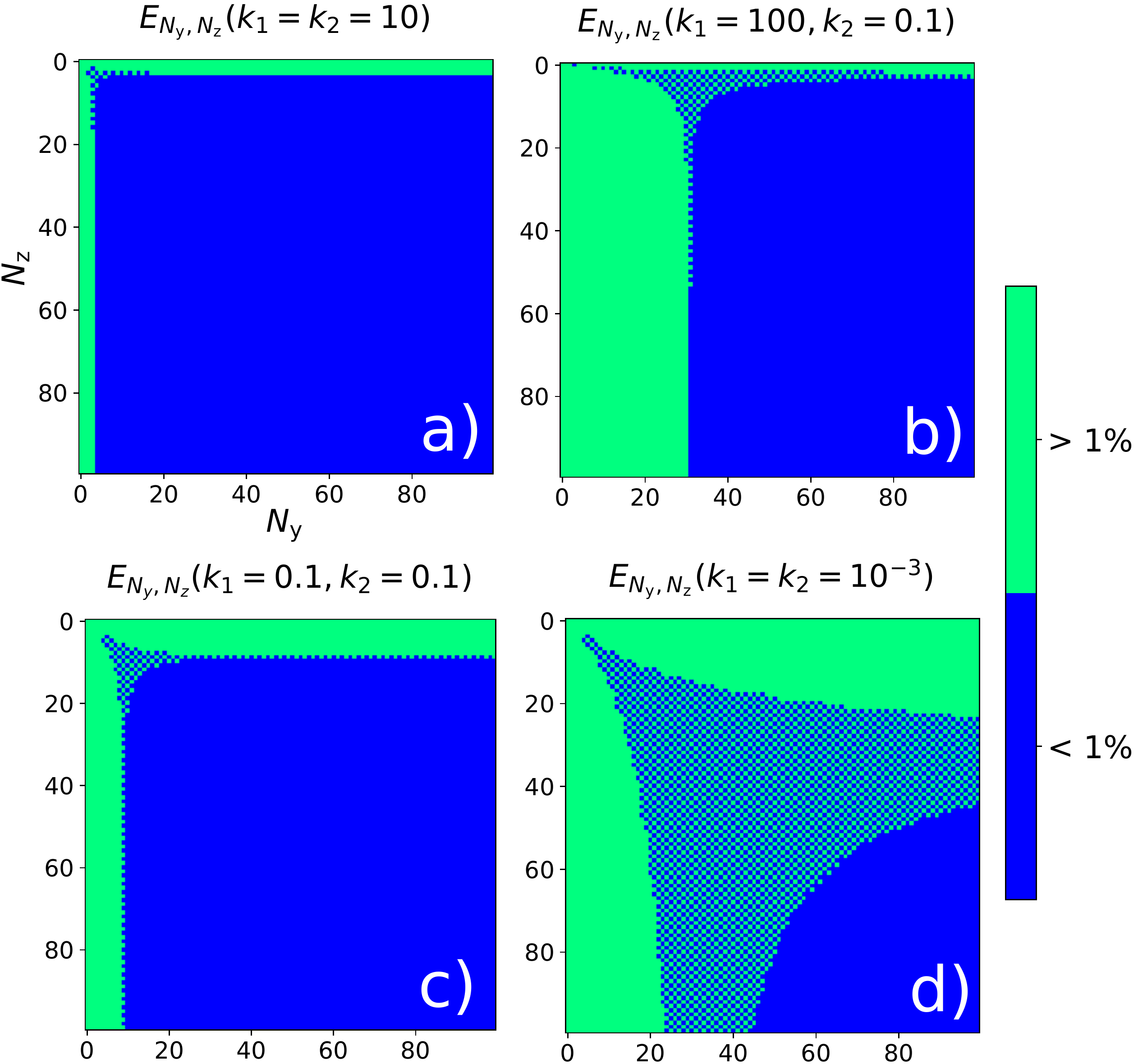}}
\caption{Relative error Eq.\ref{3DRelativeError} made in the numerical evaluation of Eq.\ref{sumConvergence} for several sets of parameters $\{k_1,k_2\}$. The worst case scenario happens for $k_1,k_2\ll 1$ shown in d).}
\label{fig:App7}       
\end{figure}

After studying the worst convergence case ($k_1,k_2 \ll 1$), the convergence of the sum in the general case for any $k_1,k_2$ values has been computed.
\begin{equation}\label{sumConvergenceApprox}
S_{N_y,N_z}(k_1,k_2)= \sum\limits_{\substack{0\leq n_y\leq N_y \\ 0\leq n_z \leq N_z}}\frac{(-1)^{n_y+n_z}\text{tanh}\left( \frac{\pi}{2}a_{n_y,n_z}\right)}{(2n_y+1)(2n_z+1)a_{n_y,n_z}} 
\end{equation}
Since no analytical form exists for the relative error in the general case (similar to Eq.\ref{limit3DRelativeError}), the relative error is computed with with the approximation $S_{\infty,\infty}(k_1,k_2)\approx S_{2000,2000}(k_1,k_2)$:
\begin{equation}
E_{N_1,N_2}(k_1,k_2) \equiv \left|\frac{S_{2000,2000}(k_1,k_2)-S_{N_1,N_2}(k_1,k_2)}{S_{2000,2000}(k_1,k_2)}\right| \label{3DRelativeError}
\end{equation}
The results are shown in Fig.\ref{fig:App7} for several $\{k_1,k_2\}$ values. As expected, the least convergent case occurs when $k_1,k_2\ll 1$ (see Fig.\ref{fig:App7} d). 

In conclusion, a sum up to $N_y,N_z=63$ is necessary to get a less than $1\%$ relative error. We went further and took into account $N_y,N_z=100$ for all the fits displayed in the main text (Fig.\ref{fig:5} and \ref{fig:6}).

\bibliographystyle{unsrt}
\bibliography{biblioTheoryEwen}

\begin{thebibliography}{10}

\bibitem{Cox2008}
Susan Cox, J.~Singleton, R.~D. McDonald, A.~Migliori, and P.~B. Littlewood.
\newblock Sliding charge-density wave in manganites.
\newblock {\em Nature Materials}, 7(1):25--30, 2008.

\bibitem{RevModPhys.60.1129}
G.~Gr\"uner.
\newblock The dynamics of charge-density waves.
\newblock {\em Rev. Mod. Phys.}, 60:1129--1181, Oct 1988.

\bibitem{Onishi_2017}
Seita Onishi, Mehdi Jamei, and Alex Zettl.
\newblock Narrowband noise study of sliding charge density waves in
  {NbSe}3nanoribbons.
\newblock {\em New Journal of Physics}, 19(2):023001, feb 2017.

\bibitem{PhysRevLett.100.096403}
D.~Le~Bolloc'h, V.~L.~R. Jacques, N.~Kirova, J.~Dumas, S.~Ravy, J.~Marcus, and
  F.~Livet.
\newblock Observation of correlations up to the micrometer scale in sliding
  charge-density waves.
\newblock {\em Phys. Rev. Lett.}, 100:096403, Mar 2008.

\bibitem{PhysRevB.85.035113}
V.~L.~R. Jacques, D.~Le~Bolloc'h, S.~Ravy, J.~Dumas, C.~V. Colin, and
  C.~Mazzoli.
\newblock Evolution of a large-periodicity soliton lattice in a current-driven
  electronic crystal.
\newblock {\em Phys. Rev. B}, 85:035113, Jan 2012.

\bibitem{PhysRevB.94.201120}
A.~Rojo-Bravo, V.~L.~R. Jacques, and D.~Le~Bolloc'h.
\newblock Collective transport of charges in charge density wave systems based
  on traveling soliton lattices.
\newblock {\em Phys. Rev. B}, 94:201120, Nov 2016.

\bibitem{Feinberg}
{Feinberg, D.} and {Friedel, J.}
\newblock Elastic and plastic deformations of charge density waves.
\newblock {\em J. Phys. France}, 49(3):485--496, 1988.

\bibitem{GORKOV1983}
L.~P GOR'KOV.
\newblock Conditions aux limites et génération des bruits périodiques par
  une onde de densité de charge.
\newblock {\em Pis’ma Zh. Eksp. Teor. Fiz.}, 1983.

\bibitem{gor1984generation}
LP~Gor’kov.
\newblock Generation of oscillations by a running charge density wave.
\newblock {\em Sov. Phys. JETP}, 59:1057--1064, 1984.

\bibitem{PhysRevLett.68.2066}
Satish Ramakrishna, Michael~P. Maher, Vinay Ambegaokar, and Ulrich Eckern.
\newblock Phase slip in charge-density-wave systems.
\newblock {\em Phys. Rev. Lett.}, 68:2066--2069, Mar 1992.

\bibitem{Gill_1986}
J~C Gill.
\newblock Thermally initiated phase-slip in the motion and relaxation of
  charge-density waves in niobium triselenide.
\newblock {\em Journal of Physics C: Solid State Physics}, 19(33):6589--6604,
  nov 1986.

\bibitem{PhysRevB.48.4860}
Ji-Min Duan.
\newblock Homogeneous quantum phase slippage in bulk charge-density-wave
  systems.
\newblock {\em Phys. Rev. B}, 48:4860--4863, Aug 1993.

\bibitem{MAKI1995313}
Kazumi Maki.
\newblock Quantum phase slip in charge and spin density waves.
\newblock {\em Physics Letters A}, 202(4):313 -- 316, 1995.

\bibitem{PhysRevB.85.241104}
A.~A. Sinchenko, P.~Lejay, and P.~Monceau.
\newblock Sliding charge-density wave in two-dimensional rare-earth tellurides.
\newblock {\em Phys. Rev. B}, 85:241104, Jun 2012.

\bibitem{PhysRevB.70.075111}
Naoki Ogawa and Kenjiro Miyano.
\newblock Optical investigation of the origin of switching conduction in
  charge-density waves.
\newblock {\em Phys. Rev. B}, 70:075111, Aug 2004.

\bibitem{PhysRevLett.117.156401}
V.~L.~R. Jacques, C.~Laulh\'e, N.~Moisan, S.~Ravy, and D.~Le~Bolloc'h.
\newblock Laser-induced charge-density-wave transient depinning in chromium.
\newblock {\em Phys. Rev. Lett.}, 117:156401, Oct 2016.

\bibitem{PhysRevB.32.2621}
M.~Prester.
\newblock Size effect in nb${\mathrm{se}}_{3}$: Length dependence of the
  threshold field.
\newblock {\em Phys. Rev. B}, 32:2621--2624, Aug 1985.

\bibitem{YETMAN1987201}
P.J. Yetman and J.C. Gill.
\newblock Size-dependent threshold fields for fröhlich conduction in niobium
  triselenide: Possible evidence for pinning by the crystal surface.
\newblock {\em Solid State Communications}, 62(3):201 -- 206, 1987.

\bibitem{MIHALY1983203}
G.~Mihály, Gy. Hutiray, and L.~Mihály.
\newblock Local distortion of pinned charge density waves in orthorombic tas3.
\newblock {\em Solid State Communications}, 48(3):203 -- 205, 1983.

\bibitem{PhysRevLett.80.5631}
H.~Requardt, F.~Ya. Nad, P.~Monceau, R.~Currat, J.~E. Lorenzo, S.~Brazovskii,
  N.~Kirova, G.~Gr\"ubel, and Ch. Vettier.
\newblock Direct observation of charge density wave current conversion by
  spatially resolved synchrotron x-ray studies in ${\mathrm{nbse}}_{3}$.
\newblock {\em Phys. Rev. Lett.}, 80:5631--5634, Jun 1998.

\bibitem{PhysRevB.61.10640}
S.~Brazovskii, N.~Kirova, H.~Requardt, F.~Ya. Nad, P.~Monceau, R.~Currat, J.~E.
  Lorenzo, G.~Gr\"ubel, and Ch. Vettier.
\newblock Plastic sliding of charge density waves: X-ray space resolved-studies
  versus theory of current conversion.
\newblock {\em Phys. Rev. B}, 61:10640--10650, Apr 2000.

\bibitem{PhysRevB.57.12781}
S.~G. Lemay, M.~C. de~Lind~van Wijngaarden, T.~L. Adelman, and R.~E. Thorne.
\newblock Spatial distribution of charge-density-wave phase slip in
  ${\mathrm{nbse}}_{3}$.
\newblock {\em Phys. Rev. B}, 57:12781--12791, May 1998.

\bibitem{PhysRevB.46.4456}
J.~McCarten, D.~A. DiCarlo, M.~P. Maher, T.~L. Adelman, and R.~E. Thorne.
\newblock Charge-density-wave pinning and finite-size effects in
  ${\mathrm{nbse}}_{3}$.
\newblock {\em Phys. Rev. B}, 46:4456--4482, Aug 1992.

\bibitem{BORODIN198673}
D.V. Borodin, F.Ya. Nad', Ya.S. Savitskaya, and S.V. Zaitsev-Zotov.
\newblock Nonlinear effects in small o-tas3 samples.
\newblock {\em Physica B+C}, 143(1):73 -- 75, 1986.

\bibitem{PhysRevB.101.125122}
E.~Bellec, I.~Gonzalez-Vallejo, V.~L.~R. Jacques, A.~A. Sinchenko, A.~P. Orlov,
  P.~Monceau, S.~J. Leake, and D.~Le~Bolloc'h.
\newblock Evidence of charge density wave transverse pinning by x-ray
  microdiffraction.
\newblock {\em Phys. Rev. B}, 101:125122, Mar 2020.

\bibitem{Batisti}
{Batisti\'{}c, I.}, {Bjelis, A.}, and {Gor'kov, L.P.}
\newblock Generation of the coherent pulses by the cdw-motion. solutions of the
  microscopic model equations.
\newblock {\em J. Phys. France}, 45(6):1049--1059, 1984.

\bibitem{hayashi2000ginzburglandau}
Masahiko Hayashi and Hideo Yoshioka.
\newblock On the ginzburg-landau free energy of charge density waves with a
  three-dimensional order, 2000.

\bibitem{PhysRevB.17.535}
H.~Fukuyama and P.~A. Lee.
\newblock Dynamics of the charge-density wave. i. impurity pinning in a single
  chain.
\newblock {\em Phys. Rev. B}, 17:535--541, Jan 1978.

\bibitem{gruner2018density}
G.~Gruner.
\newblock {\em Density Waves In Solids}.
\newblock CRC Press, 2018.

\bibitem{lee1979electric}
PA~Lee and TM~Rice.
\newblock Electric field depinning of charge density waves.
\newblock {\em Physical Review B}, 19(8):3970, 1979.

\bibitem{riley2002mathematical}
K.F. Riley, M.P. Hobson, and S.J. Bence.
\newblock {\em Mathematical Methods for Physics and Engineering: A
  Comprehensive Guide}.
\newblock Cambridge University Press, 2002.

\bibitem{PhysRevB.29.755}
A.~Zettl and G.~Gr\"uner.
\newblock Phase coherence in the current-carrying charge-density-wave state:
  ac-dc coupling experiments in nb${\mathrm{se}}_{3}$.
\newblock {\em Phys. Rev. B}, 29:755--767, Jan 1984.

\bibitem{PhysRevB.43.8421}
J.~P. Pouget, B.~Hennion, C.~Escribe-Filippini, and M.~Sato.
\newblock Neutron-scattering investigations of the kohn anomaly and of the
  phase and amplitude charge-density-wave excitations of the blue bronze
  ${\mathrm{k}}_{0.3}$${\mathrm{moo}}_{3}$.
\newblock {\em Phys. Rev. B}, 43:8421--8430, Apr 1991.

\bibitem{brun2005charge}
C~Brun, JC~Girard, ZZ~Wang, J~Marcus, J~Dumas, and C~Schlenker.
\newblock Charge-density waves in rubidium blue bronze rb 0.3 mo o 3 observed
  by scanning tunneling microscopy.
\newblock {\em Physical Review B}, 72(23):235119, 2005.

\bibitem{mallet1999contrast}
P~Mallet, KM~Zimmermann, Ph~Chevalier, J~Marcus, JY~Veuillen, and JM~Gomez
  Rodriguez.
\newblock Contrast reversal of the charge density wave stm image in purple
  potassium molybdenum bronze k 0.9 mo 6 o 17.
\newblock {\em Physical Review B}, 60(3):2122, 1999.

\bibitem{gammie1989scanning}
G~Gammie, JS~Hubacek, SL~Skala, RT~Brockenbrough, JR~Tucker, and Joseph~W
  Lyding.
\newblock Scanning tunneling microscopy of the charge-density wave in
  orthorhombic tas 3.
\newblock {\em Physical Review B}, 40(17):11965, 1989.

\bibitem{brun2010surface}
Christophe Brun, Zhao-Zhong Wang, Pierre Monceau, and Serguei Brazovskii.
\newblock Surface charge density wave phase transition in nbs e 3.
\newblock {\em Physical review letters}, 104(25):256403, 2010.

\bibitem{brun2009scanning}
Christophe Brun, Zhao-Zhong Wang, and Pierre Monceau.
\newblock Scanning tunneling microscopy at the nbse 3 surface: Evidence for
  interaction between q 1 and q 2 charge density waves in the pinned regime.
\newblock {\em Physical Review B}, 80(4):045423, 2009.

\bibitem{brazovskii2012scanning}
Serguei Brazovskii, Christophe Brun, Zhao-Zhong Wang, and Pierre Monceau.
\newblock Scanning-tunneling microscope imaging of single-electron solitons in
  a material with incommensurate charge-density waves.
\newblock {\em Physical review letters}, 108(9):096801, 2012.

\bibitem{fang2007stm}
A~Fang, N~Ru, IR~Fisher, and A~Kapitulnik.
\newblock Stm studies of tbte 3: evidence for a fully incommensurate charge
  density wave.
\newblock {\em Physical review letters}, 99(4):046401, 2007.

\bibitem{fu2016multiple}
Ling Fu, Aaron~M Kraft, Bishnu Sharma, Manoj Singh, Philip Walmsley, Ian~R
  Fisher, and Michael~C Boyer.
\newblock Multiple charge density wave states at the surface of tbt e 3.
\newblock {\em Physical Review B}, 94(20):205101, 2016.

\bibitem{burk1991charge}
B~Burk, RE~Thomson, A~Zettl, and John Clarke.
\newblock Charge-density-wave domains in 1t-tas 2 observed by satellite
  structure in scanning-tunneling-microscopy images.
\newblock {\em Physical review letters}, 66(23):3040, 1991.

\bibitem{murphy2003surface}
BM~Murphy, J~Stettner, M~Traving, M~Sprung, I~Grotkopp, M~M{\"u}ller,
  CS~Oglesby, M~Tolan, and W~Press.
\newblock Surface behaviour at the charge density wave transition in nbse2.
\newblock {\em Physica B: Condensed Matter}, 336(1-2):103--108, 2003.

\bibitem{PhysRevB.42.8791}
X-M Zhu, R.~Moret, H.~Zabel, I.~K. Robinson, E.~Vlieg, and R.~M. Fleming.
\newblock Grazing-incidence x-ray study of the charge-density-wave phase
  transition in ${\mathrm{k}}_{0.3}$${\mathrm{moo}}_{3}$.
\newblock {\em Phys. Rev. B}, 42:8791--8794, Nov 1990.

\bibitem{ru2008charge}
Nancy Ru.
\newblock {\em Charge density wave formation in rare-earth tritellurides}.
\newblock 2008.

\bibitem{schlenker1989low}
C.~Schlenker.
\newblock {\em Low-dimensional electronic properties of molybdenum bronzes and
  oxides}.
\newblock Physics and chemistry of materials with low-dimensional structures.
  Kluwer Academic Publishers, 1989.

\bibitem{belle2019}
Ewen Bellec.
\newblock {\em Study of charge density wave materials under current by X-ray
  diffraction}.
\newblock PhD thesis, 2019.
\newblock Thèse de doctorat dirigée par Le Bolloc'h, David Physique
  Université Paris-Saclay (ComUE) 2019.

\bibitem{PhysRevB.40.11965}
G.~Gammie, J.~S. Hubacek, S.~L. Skala, R.~T. Brockenbrough, J.~R. Tucker, and
  J.~W. Lyding.
\newblock Scanning tunneling microscopy of the charge-density wave in
  orthorhombic ${\mathrm{tas}}_{3}$.
\newblock {\em Phys. Rev. B}, 40:11965--11968, Dec 1989.

\bibitem{PhysRevB.72.235119}
C.~Brun, J.~C. Girard, Z.~Z. Wang, J.~Marcus, J.~Dumas, and C.~Schlenker.
\newblock Charge-density waves in rubidium blue bronze
  ${\mathrm{rb}}_{0.3}\mathrm{Mo}{\mathrm{o}}_{3}$ observed by scanning
  tunneling microscopy.
\newblock {\em Phys. Rev. B}, 72:235119, Dec 2005.

\bibitem{maki1989impurity}
Kazumi Maki and Attila Virosztek.
\newblock Impurity pinning of charge-density waves and spin-density waves.
\newblock {\em Physical Review B}, 39(13):9640, 1989.

\bibitem{ZYBTSEV201534}
S.G. Zybtsev and V.Ya. Pokrovskii.
\newblock Strain-induced formation of ultra-coherent cdw in quasi
  one-dimensional conductors.
\newblock {\em Physica B: Condensed Matter}, 460:34 -- 38, 2015.
\newblock Special Issue on Electronic Crystals (ECRYS-2014).

\bibitem{mihaly1986dielectric}
Laszlo Mihaly and GX~Tessema.
\newblock Dielectric hysteresis and relaxation in the charge-density-wave
  compound k 0.3 moo 3.
\newblock {\em Physical Review B}, 33(8):5858, 1986.

\bibitem{mihaly1984onset}
L~Mihaly and G~Gr{\"u}ner.
\newblock The onset of current carrying charge density wave state in tas3:
  Switching, hysteresis, and oscillation phenomena.
\newblock {\em Solid state communications}, 50(9):807--811, 1984.

\bibitem{zettl1982onset}
A~Zettl and G~Gr{\"u}ner.
\newblock Onset of charge-density-wave conduction: Switching and hysteresis in
  nb se 3.
\newblock {\em Physical Review B}, 26(4):2298, 1982.

\bibitem{danneau2002motional}
R~Danneau, A~Ayari, D~Rideau, H~Requardt, JE~Lorenzo, L~Ortega, P~Monceau,
  R~Currat, and G~Gr{\"u}bel.
\newblock Motional ordering of a charge-density wave in the sliding state.
\newblock {\em Physical review letters}, 89(10):106404, 2002.

\bibitem{peyrard2012physique}
Michel Peyrard.
\newblock {\em Physique des solitons}.
\newblock EDP Sciences, 2012.

\bibitem{Laplacian}
F.~Pockels.
\newblock Über die partielle differentialgleichung und deren auftreten in die
  mathematischen physik.
\newblock {\em Z. Math. Physik}, 37:100--105, 1892.

\bibitem{Lerch}
M.~Lerch.
\newblock Note sur la fonction $\mathfrak{K}(w,x,s) = \sum_{k=0}^\infty
  \frac{e^{2 k \pi i x}}{(w+k)^s}$.
\newblock {\em Acta Math.}, 11(1-4):19–24, 1887.

\bibitem{Erdelyi}
Arthur Erd{\'e}lyi, Wilhelm Magnus, Fritz Oberhettinger, and Francesco~G.
  Tricomi.
\newblock {\em Higher Transcendental Functions. {V}ol. {I}}.
\newblock McGraw-Hill Book Company, Inc., New York-Toronto-London, 1953.
\newblock Reprinted by Robert E. Krieger Publishing Co. Inc., 1981.

\end{thebibliography}
%
%
%

\end{document}